\documentclass[12pt, draftclsnofoot, journal, letter, onecolumn]{IEEEtran}
\usepackage{graphicx}
\usepackage{epsfig}
\usepackage{latexsym}
\usepackage{amsfonts}
\usepackage{here}
\usepackage{rawfonts}
\usepackage[latin1]{inputenc}
\usepackage[T1]{fontenc}
\usepackage{calc}
\usepackage{url}
\usepackage{enumerate}
\usepackage{color}
\usepackage[tbtags]{amsmath}
\usepackage{amssymb}
\usepackage{upref}
\usepackage{epic,eepic}
\usepackage{times}
\usepackage{dsfont}
\usepackage{comment}
\usepackage{cite}
\usepackage{graphicx}
\usepackage[font={small}]{caption}
\usepackage[font={small}]{subcaption}

\usepackage{stfloats}
\usepackage{mathtools}
\usepackage{amsmath}

\newtheorem{theorem}{\bf{Theorem}}

\newtheorem{corollary}{\bf{Corollary}}
\newtheorem{lemma}{\bf{Lemma}}
\newtheorem{remark}{\bf{Remark}}

\ifCLASSINFOpdf
\else
\fi

\begin{document}

\title{Stochastic Design and Analysis of \\Wireless Cloud Caching Networks}

\author{Seyed~Mohammad~Azimi-Abarghouyi,\\Masoumeh Nasiri-Kenari,~\IEEEmembership{Senior~Member,~IEEE}, and~M\' erouane~Debbah,~\IEEEmembership{Fellow,~IEEE}
\thanks{S.M. Azimi-Abarghouyi and M. Nasiri-Kenari are with the Department of Electrical Engineering, Sharif
University of Technology, Tehran, Iran, (Emails: azimi$\_$sm@ee.sharif.edu; mnasiri@sharif.edu). M. Debbah is with the Large Networks and Systems Group, CentraleSup\' elec, Universit\' e Paris-Saclay, Gif-sur-Yvette, France, and also with the Mathematical and Algorithmic Sciences Lab, Huawei France R\&D, Paris, France, (Email: merouane.debbah@centralesupelec.fr).}
}


\maketitle

\vspace{0pt}
\begin{abstract}
This paper develops a stochastic geometry-based approach for the modeling, analysis, and optimization of wireless cloud caching networks comprised of multiple-antenna radio units (RUs) inside clouds. We consider the Matern cluster process to model RUs and the probabilistic content placement to cache files in RUs. Accordingly, we study the exact hit probability for a user of interest for two strategies; closest selection, where the user is served by the closest RU that has its requested file, and best selection, where the serving RU having the requested file provides the maximum instantaneous received power at the user. As key steps for the analyses, the Laplace transform of out of cloud interference, the desired link distance distribution in the closest selection, and the desired link received power distribution in the best selection are derived. Also, we approximate the derived exact hit probabilities for both the closest and the best selections in such a way that the related objective functions for the content caching design of the network can lead to tractable concave optimization problems. Solving the optimization problems, we propose algorithms to efficiently find their optimal content placements. Finally, we investigate the impact of different parameters such as the number of antennas and the cache memory size on the caching performance.

\end{abstract}


\section{Introduction}
The global mobile data traffic has been experiencing an explosive growth of content-oriented services such as social networking, video streaming, and smartphone application downloads. Therefore, the key challenge for the future network design is to enable low-latency massive content delivery from content providers to end users [1]. Called edge caching, an emerging key solution is caching popular files during off-peak times at the network edge, such as small cells and helper nodes, which mitigates the severe peak-hour traffic in backhaul links and improves the user quality of service [2]-[5]. In general, caching networks operate in two distinct phases [6]. The first phase is "cache placement", where based on statistics of user requests and a caching strategy each cache-enabled node is to place files in its cache that may be limited by a memory size. The second phase is "content delivery", where files are delivered according to the cache state once actual user requests are revealed. One commonly used caching strategy is the probabilistic content placement, also referred to as geographic caching [7] and independent random caching [8] in the literature, where a complete file may be stored in a cache with a fixed probability. Recently, a new caching strategy called coded caching is proposed where each file is partitioned into several segments and caching different segments in different caches can more enhance the caching gain [6]. On the other hand, the increasing randomness and irregularity in the locations of nodes in a wireless network has led to a growing interest in the use of stochastic geometry and Poisson point processes (PPPs) for flexible and tractable spatial network modeling and analysis [9]-[11]. Modeling the locations of nodes in caching networks by the homogeneous Poisson point process (HPPP) [12, Def. 2.8] and using the probabilistic content placement, the optimal tier-level content placement for non-cooperative and cooperative heterogeneous cellular networks are studied in [13] and [14], respectively, the optimal cache hit probability and the optimal throughput content placement for device-to-device (D2D) networks are studied in [15], and the D2D optimal density of successful receptions is studied in [16]. Furthermore, the performance of small cell networks in terms of the outage probability and the average delivery rate is evaluated in [17], and the effect of retransmissions on the optimal caching in small cell networks is investigated in [18]. In [19] and [20], with the HPPP, coded caching strategies are used and their coded cache placements are investigated. A combined coded and uncoded strategy for the content caching placement in cooperative small cell networks modeled by the HPPP is also proposed in [21].

As another promising solution to meet the explosively growing data demand, cloud radio access network (C-RAN) architectures with ultra-dense radio units (RUs) have recently been proposed in [22]. C-RANs are not only to incorporate cloud computing into radio access networks via cloud processors (CPs) but also to reduce system cost in comparison to other base station (BS) deployments. Using coordinated multi-point (CoMP) transmissions, C-RANs are also able to alleviate severe interference in ultra-dense networks. However, their main weak points are significant computation complexity of centralized processings in CPs and aggregated traffic due to capacity limitations of fronthaul links [23]-[24]. Also, the HPPP cannot model the emerging user-centric deployments of C-RANs, where RUs may be deployed at places with high user density [25]. In such deployments, it is important to take into account non-uniformity as well as the correlation that may exist between the locations. Accordingly, the Third Generation Partnership Project (3GPP) has considered user-centric models in [26]-[27]. Also, user-centric models based on Poisson cluster processes (PCPs), namely Thomas cluster process (TCP) [12, Def. 3.5] and Matern cluster process (MCP) [12, Def. 3.6], have recently been studied for different wireless networks [28]-[30]. 

In this paper, inspired by the edge caching and C-RAN architectures, we propose wireless cloud caching networks for user-centric applications, where RUs inside clouds are located based on an MCP and there is no CP requirements. According to a popularity distribution, we use the probabilistic content placement to cache files in RUs having multiple antennas and a limited memory size, and we use a low-complexity zero-forcing transmission strategy to distributedly coordinate the RUs of each cloud and suppress its intra cloud interference. Accordingly, we propose two selection strategies, the closest selection and the best selection, to select a RU to service a user in the network and derive their exact hit probabilities at any location of a user of interest inside a representative cloud. As key steps in the analyses, the Laplace transform (LT) of the out of cloud interference is characterized. We further characterize the distributions of the desired signal power in the best selection and the desired link distance in the closest selection. We also approximate the exact hit probabilities for both the closest and the best selections with the results that are convenient for a caching design over the network. Then, we propose optimization problems to find the content placement probabilities for both selection strategies that maximize the hit probabilities. We prove that the optimization problems are concave. Also, to solve them, we propose efficient optimization algorithms.

We investigate the impact of different parameters of the system model on the performance in terms of the hit probability. Our analysis reveals that a higher number of antennas or distance of the user of interest to the center of the representative cloud has a degrading effect on the performance. Also, increasing
the memory size or the skewness parameter of the popularity distribution increases the hit probability. As another observation, there exists a pathloss exponent that can provide the highest hit probability. Also, while the best selection achieves a higher performance than the closest selection, the proposed probabilistic content placement significantly outperforms a benchmark scheme that caches only the most popular files.

The rest of the paper is organized as follows. Section II describes the system model including the spatial model, the caching strategy, and the performance metric. Section III introduces the transmission strategy as well as the selection strategies and characterizes their exact and approximate hit probabilities. Section IV presents the content caching design by optimizing the approximate hit probability results. Section V presents the numerical and simulation results. Finally, Section VI concludes the paper.

\section{System Model}
In this section, we describe a mathematical model for wireless cloud caching networks as shown in Fig. 1 and define its performance
metric. 

\subsection{Spatial Setup and Signal Model}

We consider a cloud architecture as a union of independently working clouds. Referred to as a cloud, RUs inside a circular region with radius $D$ cooperatively serve different users inside the region with a CoMP transmission strategy. In addition, each cloud's RUs can interfere on the users that belong to other clouds.
\begin{table*}[t]
\caption {Summary of Notation} 
\vspace{-10pt}
\begin{center}
\resizebox{16cm}{!} {
    \begin{tabular}{ l | l }
  
   \hline
    \hline
    \textbf{Notation} & \centerline{\textbf{Description}} \\ \hline
    $\Phi^\text{U}$ & Spatial point process of users\\ \hline
    $\Phi$ & Spatial point process of RUs\\ \hline
    $\mathbf{x}$; $\mathbf{x}_\text{o}$ & Center of a cloud; center of the representative cloud \\ \hline
    $\Phi_{\mathbf{x}}$ & Spatial point process of the RUs in the cloud with center $\mathbf{x}$\\ \hline
    $\Phi_{\mathbf{x}_\text{o}}$ & Spatial point process of the RUs in the representative cloud\\ \hline
    $\lambda$ & Intensity of RUs in each cloud  \\ \hline
    $\Phi_{\rm p}$; $\lambda_{\rm p}$ & Parent point process of cloud centers; intensity of $\Phi_{\rm p}$ \\ \hline
    $\alpha_\text{i}$; $\alpha_\text{o}$ & Path loss exponent from $\Phi_{\mathbf{x}_\text{o}}$; path loss exponent from $\Phi_{\mathbf{x}}$\\ \hline
    $P$; $\beta$; $\sigma^2$ & Transmit power; SINR threshold; noise power\\ \hline
    $D$ & Radius of each cloud region \\ \hline
    $d_\text{g}$ & Radius of guard disk \\ \hline
     $M$ & Number of antennas of each RU  \\ \hline
     $f_i^{\mathbf{x}}$ & $i$-th popular file in the cloud $\Phi_{\mathbf{x}}$\\ \hline
     $q_i^{\mathbf{x}}$ & Request probability of $i$-th popular file in the cloud $\Phi_{\mathbf{x}}$\\ \hline
     $p_i^{\mathbf{x}}$ & Cache probability of $i$-th popular file in the cloud $\Phi_{\mathbf{x}}$\\ \hline
     $\gamma_{\mathbf{x}}$ & Skewness parameter of the popularity distribution in the cloud $\Phi_{\mathbf{x}}$\\ \hline
     $N_{\mathbf{x}}$ & Library size of files in the cloud $\Phi_{\mathbf{x}}$\\ \hline
     $N_{\text{c}}^{\mathbf{x}}$ & Memory size of caches in the cloud $\Phi_{\mathbf{x}}$\\ \hline
     ${\cal A}^{\mathbf{x}}$ & Set of Active RUs (in the channel of interest) in the cloud $\Phi_{\mathbf{x}}$\\ \hline
    \hline
    \end{tabular}}
 
\end{center}
\vspace{-15pt}
\end{table*}

We consider a user-centric spatial modeling for the locations of RUs and users over the infinite region, i.e., $\mathbb{R}^2$. The RUs are distributed as an MCP [12] ${\Phi}$, defined as a union of offspring points that are located around parent points. The parent point process is modeled as an HPPP ${\Phi}_\text{p}$ with intensity $\lambda_\text{p}$ and denotes the cloud centers. The offspring points of the cloud with the center $\mathbf{x}\in\Phi_{\rm p}$ form a finite HPPP (FHPPP) [31] ${\Phi}_\mathbf{x}$ with intensity $\lambda$ over the disk $\mathbf{b}(\mathbf{x},D)$ as the cloud region, such that ${\Phi} = \cup_{\mathbf{x} \in \Phi_\text{p}}{\Phi}_\mathbf{x}$.\footnote{In our model, clouds can overlap with each other. However, each cloud is independently working from other clouds in overlapping areas to service its users due to the limited capability of cloud processors and the disjoint nature of cloud content interests and service providers.} The users are uniformly and independently distributed inside clouds according to an MCP $\Phi^\text{U}$ conditionally independent of $\Phi$ given the parent points. For the parent point $\mathbf{x} \in \Phi_\text{p}$, the user offspring point process is denoted by $\Phi_\mathbf{x}^\text{U}$. As the worst-case scenario, we assume that the number of users is much higher than the number of RUs in a cloud. The user-centric model can incorporate correlation across the locations of users and RUs that exists due to the deployment of RUs at the places of high user density. This model is suitable for various use-case scenarios as: i) cloud access networks, where different types of restricted regions, such as libraries, shopping malls, campuses, and bars, have access points as RUs to meet their users' demands, ii) cloud small-cell BSs, where small-cell BSs as RUs are located at dense areas of people over a city, and iii) cloud BSs, where a BS is formed by a number of distributed antenna terminals as RUs.

We assume that each RU is equipped with $M$ antennas and transmits at the same power $P$. On the other hand, users have single antennas. Also, each RU is assumed to be fully-loaded by several users and serves them in different channels.

In the proposed setup, a RU located at $\mathbf{z}\in {\Phi}$ transmits information $s_\mathbf{z}$ through a linear beamforming vector $\mathbf{v}_\mathbf{z}$ with unit norm, i.e., $\|\mathbf{v}_\mathbf{z}\| =1$. With no loss of generality, we focus on the design of a representative cloud $\Phi_{\mathbf{x}_\text{o}}$ with intensity $\lambda$ of RUs over the disk $\mathbf{b}(\mathbf{x}_\text{o},D)$. And, we conduct the performance analysis for a user of interest located at the origin $\mathbf{o}$ inside the representative cloud and assigned to a channel of interest, such that $\|\mathbf{x}_\text{o}\|\leq D$ is the distance from the user of interest to the center of the representative cloud.\footnote{The location of the user of interest can be anywhere in $\mathbf{b}(\mathbf{x}_\text{o},D)$. The origin $\mathbf{o}$ and ${\mathbf{x}_\text{o}}$ are relatively determined in a coordinate system.} Further, to mitigate severe interference from close interfering RUs, we consider a guard disk with radius $d_\text{g}$ centered at the user of interest, i.e., $\mathbf{b}(\mathbf{o},d_\text{g})$, inside which the RUs of clouds except the representative cloud are not allowed to work in the channel of interest. Then, denoting the sets of RUs working in the channel of interest, referred to as active RUs, in ${\Phi}_{\mathbf{x}_\text{o}}$ and ${\Phi}_{\mathbf{x}}$ with ${\cal A}_{\mathbf{x}_\text{o}}$ and ${\cal A}_{\mathbf{x}}$, respectively, the received signal at the user of interest is given by
\begin{eqnarray}
{y_\mathbf{o}} = \mathop \sum \limits_{\bf{z} \in {{{\cal A}_{\mathbf{x}_\text{o}}}}} {\| \mathbf{z} \|^{ - \frac{\alpha_\text{i} }{2}}}{{\bf{h}}_\mathbf{zo}^{*}}{{\mathbf{v}}_\mathbf{z}}{s_\mathbf{z}} +\mathop \sum \limits_{\mathbf{x} \in {\Phi_\text{p}}}\mathop \sum \limits_{\bf{z} \in {{\cal A}_{\mathbf{x}}}} {\| \mathbf{z} \|^{ - \frac{\alpha_\text{o} }{2}}}{{\bf{h}}_\mathbf{zo}^{*}}{{\mathbf{v}}_\mathbf{z}}{s_\mathbf{z}} + n_\mathbf{o},
\end{eqnarray}
where $*$ denotes the hermitian transpose and $\mathbf{h}_\mathbf{zo}=  \small{\begin{bmatrix}
           h_\mathbf{zo}^1 \\
           \vdots \\
          h_\mathbf{zo}^M
         \end{bmatrix}}$ is the channel vector between the RU at $\mathbf{z}$ and the user of interest, whose entries are independent and identically distributed i.i.d. quasi-static complex Gaussian random variables with zero mean and unit variance, i.e., ${\cal {CN}}(0,1)$. Also, $\alpha_\text{i}$ and $\alpha_\text{o}$ represent the pathloss exponents from the RUs of the representative cloud and the RUs of other clouds, respectively, such that $\alpha_\text{i}<\alpha_\text{o}$,\footnote{This is due to the fact that the RUs of other clouds are usually outside the region of the representative cloud and far away from the user of interest.} and $n_\mathbf{o}$ denotes the additive Gaussian noise with zero mean and variance $\sigma^2$, i.e., ${\cal {CN}}(0,\sigma^2)$. 
\begin{figure}[tb!]
\centering

\includegraphics[width =4.5in]{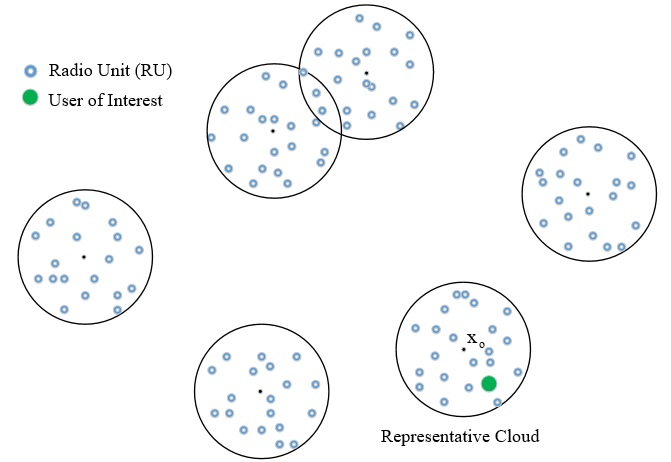}

\caption{An illustration of the system model for a finite piece of wireless cloud caching networks.}
\vspace{-10pt}
\end{figure}

\subsection{Content Caching Strategy}
We consider that clouds may have different contents.\footnote{While the tools developed in this paper can be extended to handle the case where clouds have shared contents and a user can be served by every clouds, it is not in the scope of this paper and is left as a promising future work.} For example, a university campus is mostly interested in science-related contents in contrast with a sports bar where the nature of popular contents is often completely different. Further, we consider a finite content library ${\cal F}^{\mathbf{x}} = [f_1^{\mathbf{x}},...,f_{N_\mathbf{x}}^{\mathbf{x}}]$ for each cloud $\Phi_\mathbf{x}, \forall \mathbf{x} \in \Phi_\text{p}$, where $N_\mathbf{x}$ is the library size and $f_i^{\mathbf{x}}$ is the $i$-th most popular file with normalized size 1. Each user of the cloud $\Phi_\mathbf{x}$ makes an independent request for each file $f_i^{\mathbf{x}}$ with probability $q_i^{\mathbf{x}}$. With no loss of generality, the order $q_1^{\mathbf{x}} \geq \ldots \geq q_{N_\mathbf{x}}^{\mathbf{x}}$ is assumed for the files. Also, for the purpose of evaluations, we will focus on the Zipf distribution [32], which is widely used for video popularity and given by
\begin{eqnarray}
q_i^{\mathbf{x}} = \frac{{{i^{ - \gamma_{\mathbf{x}} }}}}{{\mathop \sum \nolimits_{j = 1}^{N_{\mathbf{x}}} {j^{ - \gamma_{\mathbf{x}} }}}},
\end{eqnarray}
where $\gamma_{\mathbf{x}}$ is the shape parameter that determines the skewness of the popularity distribution in the cloud region $\mathbf{b}(\mathbf{x},D)$. In general, the content popularity distribution can be characterized by machine learning tools [33]-[34].

We assume that the RUs of $\Phi_\mathbf{x}$ have a local cache with memory size $N_\text{c}^\mathbf{x}<N_\mathbf{x}$ in which they can store $N_\text{c}^\mathbf{x}$ files that may be of interest to the users. We apply the probabilistic content placement to randomly select files for caching at different RUs of $\Phi_\mathbf{x}$. In this way, the $i$-th file is cached with a fixed probability $p_i^\mathbf{x}$, referred to as "cache probability". Thus, due to the independent thinning, the distribution of RUs having $f_i^\mathbf{x}$ follows an FHPPP ${\Phi}_\mathbf{x}^i$ with intensity $p_i^{\mathbf{x}} \lambda$ over the disk $\mathbf{b}(\mathbf{x},D)$. Also, due to the cache memory size, we have the constraint $\sum\limits_{i = 1}^{N_\mathbf{x}} p_i^\mathbf{x} \le N_\text{c}^\mathbf{x}$. We further define the cache probability set ${\cal P}^\mathbf{x} = \left\{p_1^\mathbf{x},...,p_N^\mathbf{x}\right\}$.
\subsection{Performance Metric}
The metric of interest for performance analysis is the hit probability, defined as the probability that the user of interest is able to successfully deliver its requested content from the representative cloud [7], [35]. In fact, a cache hit event happens when the two following cases are satisfied: (i) the content of interest is available in the cache of at least one RU in the cloud, and (ii) the user of interest, according to the transmission strategy, is in the coverage of RUs that have this content. As defined, the hit probability is able to quantify the reduced fraction of load and mean latency of the backhaul network [13]. 

Lets define $\text{P}_{i,\mathbf{x}_\text{o}}^\text{hit}$ as the conditional hit probability in the case that the user of interest requests the $i$-th most popular file of the representative cloud, i.e., $f_i^{\mathbf{x}_\text{o}}$. Then, for the purpose of analysis, $\text{P}_{i,\mathbf{x}_\text{o}}^\text{hit} = \mathbb{P}(\text{SINR}_{i,\mathbf{x}_\text{o}} > \beta) $, where $\text{SINR}_{i,\mathbf{x}_\text{o}}$ is the signal-to-interference-plus-noise ratio (SINR) of the transmission strategy in delivering $f_i^{\mathbf{x}_\text{o}}$ from the RUs of $\Phi_{\mathbf{x}_\text{o}}$ and $\beta$ is the minimum required threshold for a successful connection. Then, the hit probability is given by
\begin{eqnarray}
\text{P}_{\mathbf{x}_\text{o}}^\text{hit} = \mathop \sum \limits_{i = 1}^{N_{\mathbf{x}_\text{o}}} {q_i^{\mathbf{x}_\text{o}}}{\text{P}_{i,\mathbf{x}_\text{o}}^\text{hit}}.
\end{eqnarray}

In the next section, we present the transmission strategy and derive its conditional hit probability for each file, i.e., ${\text{P}_{i,\mathbf{x}_\text{o}}^\text{hit}}, \forall i$, which is required for the evaluation of the hit probability in (3). In Section IV, we discuss how to find the cache probabilities, i.e., ${\cal P}^{\mathbf{x}_\mathbf{o}}$, to maximize the hit probability. 
\section{Transmission Strategy and its Hit Proability Analysis}
As the CoMP transmission strategy, we use a zero-forcing transmission in each cloud that can nullify intra-cloud interference from at most $M-1$ RUs while maximizing the desired signal power at the user of interest, as also used for the beamforming scheme in [36]. This decentralized strategy is suitable for low-latency cloud architectures with low-complexity edge processing capabilities. To do so, the serving RU located at $\mathbf{z} \in {\Phi}_{\mathbf{x}_\text{o}}$ selects beamforming vector $\mathbf{v}_\mathbf{z}$ solving
\begin{eqnarray}
\mathop {\max }\limits_{{\mathbf{v}_\mathbf{z}}} \left| {{\mathbf{h}_\mathbf{zo}^{*}}{\mathbf{v}_\mathbf{z}}} \right|^2
\end{eqnarray}
\hspace{+170pt}subject to:
\begin{eqnarray}
\hspace{+75pt}\left\{\begin{matrix}
{{\mathbf{h}_\mathbf{zy}^{*}}{\mathbf{v}_\mathbf{z}}} = 0 \hspace{+10pt} \text{for} \hspace{+5pt} \mathbf{y} \in {\cal A}_{{\mathbf{x}_\text{o}}}^{\text{U}} \backslash \mathbf{o}, \\
\| {{\mathbf{v}_\mathbf{z}}} \| = 1. \hspace{+85pt} \nonumber
\end{matrix}\right.
\end{eqnarray}
where ${\cal A}_{{\mathbf{x}_\text{o}}}^{\text{U}} \subset {\Phi}_{\mathbf{x}_\text{o}}^\text{U}$ is the set of users of the representative cloud that are allocated to the channel of interest. This design always exists when $M \geq n({\cal A}_{{\mathbf{x}_\text{o}}}^{\text{U}})$, where $n(\cdot)$ denotes the number of elements in a set. Since each RU serves only one user in a channel, if $M < n(\Phi_{\mathbf{x}_\text{o}})$, only $M$ RUs can be active in the channel of interest and the remaining RUs in $\Phi_{\mathbf{x}_\text{o}}$ cooperate indirectly by transmitting over other channels. Then, this transmission strategy constrains $n({\cal A}_{\mathbf{x}_\text{o}}) = \min\left\{M,n(\Phi_{\mathbf{x}_\text{o}})\right\}$ and $n({\cal A}_{{\mathbf{x}_\text{o}}}^{\text{U}}) = n({\cal A}_{\mathbf{x}_\text{o}})$. We assume that the same transmission strategy is used in other clouds $\Phi_{\mathbf{x}}, \forall \mathbf{x} \in \Phi_\text{p}$, in delivering their files to their users. Also, assuming downlink-uplink channel reciprocity, we consider that each RU can learn its downlink channels to the users of its cloud by means of uplink orthogonal pilot training symbols.

Then, from (1), the SINR of the user of interest at the origin in delivering the requested file $f_i^{\mathbf{x}_\text{o}}$ can be expressed as
\begin{eqnarray}
\text{SINR}_{i,\mathbf{x}_\text{o}} =\frac{{P{|\mathbf{h}_\mathbf{zo}^{*}\mathbf{v}_\mathbf{z}|^2}{{\| \mathbf{z} \|}^{-\alpha_\text{i}}}}}{{{\sigma ^2} + {\cal I_\text{o}}}},
\end{eqnarray}
where $\mathbf{z}$ is the location of the serving RU among the RUs in the representative cloud that have $f_i^{\mathbf{x}_\text{o}}$, i.e., $\Phi_{\mathbf{x}_\text{o}}^i$, and ${\cal I_\text{o}} = \mathop \sum_{\mathbf{x} \in {\Phi}_{\text{p}}}\sum_{\mathbf{y}\in {\cal A}_{\mathbf{x}}} P{|\mathbf{h}_\mathbf{yo}^{*} \mathbf{v}_\mathbf{y}|^2}{{\| \mathbf{y} \|}^{-\alpha_\text{o}}}$ denotes the interference from other clouds, referred to as "out of cloud interference". For notational simplicity, let us define $g_z = |\mathbf{h}_\mathbf{zo}^{*}\mathbf{v}_\mathbf{z}|^2$ and $f_{y} = |\mathbf{h}_\mathbf{yo}^{*} \mathbf{v}_\mathbf{y}|^2$. It is shown in [37, Appendix A] that $g_z$ given $l$ active RUs in the representative cloud has a chi-squared distribution with $2(M-l+1)$ degrees of freedom and $f_y$ has exponential distribution with unit-mean. 

We consider two RU-selection strategies for the serving RU:\footnote{In the case that several users select an RU according to the selection strategies, the RU allocates different channels to serve all of them.}

\textbf{1) Closest Selection.} Here, as the highest long-term received power selection, the user of interest is served by the RU that has the requested file and also provides maximum received power averaged over small-scale fading. In our model, this leads to
\begin{eqnarray}
{{\rm{\mathbf{z}}_\text{CS}}} = \arg \mathop {\min }\limits_{{\mathbf{y}} \in \left\{{\Phi}_{\mathbf{x}_\text{o}}^i\mid n({\Phi}_{\mathbf{x}_\text{o}}^i)>0 \right\}} \| {{\mathbf{y}}} \|.
\end{eqnarray}
It is clear that this strategy implies that a user is served by the RU whose Voronoi cell the user resides in, considering only the RUs that have the requested file.

\textbf{2) Best Selection.}\footnote{The name "best selection" does not refer that this selection strategy has the maximum caching performance over all possible selection strategies. It refers to the best RU in providing the maximum received power.} Here, as the highest instantaneous received power selection, the serving RU has the maximum power among the RUs having the requested file at the user of interest, including small-scale fading. This leads to
\begin{eqnarray}
{{\rm{\mathbf{z}}_\text{BS}}} = \arg \mathop {\max }\limits_{{\mathbf{y}} \in \left\{{\Phi}_{\mathbf{x}_\text{o}}^i\mid n({\Phi}_{\mathbf{x}_\text{o}}^i)>0 \right\}} P g_y\| {{\mathbf{y}}} \|^{-{\alpha_\text{i}}}.
\end{eqnarray}
The handover rate of the best selection is dependent on the doppler-spread of channels and hence suitable for applications where relatively fast enough updating of RU connections is endurable. Hence, cloud access networks over restricted regions that experience slow fadings are among its highly promising applications. Such selection can be considered as a type of fast cloud processing in C-RANs [38]-[39]. It is also notable that this selection can be
simply implemented in networks since the instantaneous power can be measured in practice.

In the following subsections, we derive the conditional hit probability for both the closest and best selection strategies.

\subsection{Closest Selection}

According to (6), let us first present a lemma which gives the distance distribution of each RU in $\Phi_\mathbf{x}, \forall \mathbf{x} \in \Phi_\text{p}$, to the user of interest. 
\begin{lemma}
The probability density function (PDF) of the distance of each RU in ${\Phi}_{\mathbf{x}}$ to the origin is a function of $\|\mathbf{x}\|$ and given by [30]
\begin{eqnarray}
f^{\|\mathbf{x}\|}(y) = \left\{ {\begin{array}{*{20}{c}}
{\frac{ {2y}}{{ D^2}} \hspace{+90pt} 0\leq y<D-\|\mathbf{x}\|,}\\
{\frac{{1}}{\pi D^2} \frac{{\partial {\cal B}_{\|\mathbf{x}\|}(y) }}{{\partial y}} \hspace{+10pt} D-\|\mathbf{x}\| \leq y<D+\|\mathbf{x}\|,}\\
{0 \hspace{+119pt} y\ge D+\|\mathbf{x}\|,}
\end{array}} \right. 
\end{eqnarray}
if $\|\mathbf{x}\|<D$, and
\begin{eqnarray}
f^{\|\mathbf{x}\|}(y) = \left\{ {\begin{array}{*{20}{c}}
{0 \hspace{+98pt} 0\leq y<\|\mathbf{x}\| -D,}\\
{\frac{{1}}{\pi D^2} \frac{{\partial {\cal B}_{\|\mathbf{x}\|}(y) }}{{\partial y}} \hspace{+10pt} \|\mathbf{x}\|-D \leq y<D+\|\mathbf{x}\|,}\\
{0 \hspace{+119.6pt} y\ge D+\|\mathbf{x}\|,}
\end{array}} \right. 
\end{eqnarray}
if $\|\mathbf{x}\|>D$, where the function ${{\cal B}_{\|\mathbf{x}\|}}$ on $|\|\mathbf{x}\|-D| \leq y<D+\|\mathbf{x}\|$ is defined as
\begin{eqnarray}
{{\cal B}_{\|\mathbf{x}\|}}(y)= D^2{\cos ^{ - 1}}\left( {\frac{{D^2 + {\|\mathbf{x}\|^2} - y^2}}{{2\|\mathbf{x}\|{D}}}} \right) + y^2{\cos ^{ - 1}}\left( {\frac{{y^2 + {\|\mathbf{x}\|^2} - D^2}}{{2\|\mathbf{x}\|{y}}}} \right) \nonumber\\- \frac{1}{2}\sqrt {\left[ {{{\left( {{y} + \|\mathbf{x}\|} \right)}^2} - D^2} \right]\left[ {D^2 - {{\left( {{y} - \|\mathbf{x}\|} \right)}^2}} \right]} .
\end{eqnarray}
\hspace{460pt}\IEEEQEDclosed
\end{lemma}

Using Lemma 1 and order statistics, conditioned on the event that there are $n$ RUs in ${\Phi}_{\mathbf{x}_\text{o}}^i$, the PDF of the serving distance from the user of interest to its closest RU, i.e., $\|\mathbf{z}_{\text{CS}}\|$, is obtained as ${f_{\|\mathbf{z}_{\text{CS}}\|}}\left( r\mid n\right) =nf^{\|\mathbf{x}_\text{o}\|}(r)(1-F^{\|\mathbf{x}_\text{o}\|}(r))^{n-1}$, where $F^{\|\mathbf{x}_\text{o}\|}$ is the cumulative distribution function (CDF) of $f^{\|\mathbf{x}_\text{o}\|}$ given in (8). Therefore, we obtain
\begin{eqnarray}
{f_{\|\mathbf{z}_{\text{CS}}\|}}\left( r\mid n\right) = \left\{ {\begin{array}{*{20}{c}}
{\frac{{2nr}}{{D^2}}\left(1-\frac{r^2}{D^2}\right)^{n-1} \hspace{+119pt} 0\leq r<D-{\|\mathbf{x}_\text{o}\|},}\\
{\frac{{n}}{\pi D^2} \frac{{\partial {\cal B}_{\|\mathbf{x}_\text{o}\|}(r)}}{{\partial r}}\left(1-\frac{{{\cal B}_{\|\mathbf{x}_\text{o}\|}(r)}}{\pi D^2}\right)^{n-1}\hspace{+13pt} D-{\|\mathbf{x}_\text{o}\|} \leq r<D+{\|\mathbf{x}_\text{o}\|},}\\
{0 \hspace{+221pt} r\ge D+{\|\mathbf{x}_\text{o}\|}.}
\end{array}} \right. 
\end{eqnarray}

Using (11), we can calculate the conditional hit probability of the user of interest for the closest selection as follows. \newline By conditioning on the number of RUs in ${\Phi}_{\mathbf{x}_\text{o}}^i$ and ${\Phi}_{\mathbf{x}_\text{o}} \backslash {\Phi}_{\mathbf{x}_\text{o}}^i$ that are two independent point processes with intensities $\lambda p_i^{\mathbf{x}_\text{o}}$ and $\lambda (1-p_i^{\mathbf{x}_\text{o}})$, respectively, we have
\begin{align}
\text{P}_{i,\mathbf{x}_\text{o}}^{\text{hit},\text{CS}} &= \mathbb{P}(\text{SINR}_{i,\mathbf{x}_\text{o}} > \beta) \nonumber\\&=\sum_{k=0}^{\infty} \sum_{t=0}^{\infty}\mathbb{P}(\text{SINR}_{i,\mathbf{x}_\text{o}} > \beta \mid n({\Phi}_{\mathbf{x}_\text{o}}^i) = k, n({\Phi}_{\mathbf{x}_\text{o}} \backslash {\Phi}_{\mathbf{x}_\text{o}}^i)= t) \mathbb{P}(n({\Phi}_{\mathbf{x}_\text{o}}^i) = k)\mathbb{P}(n({\Phi}_{\mathbf{x}_\text{o}} \backslash {\Phi}_{\mathbf{x}_\text{o}}^i)= t)\nonumber\\&\stackrel{(a)}{=} \sum_{k=1}^{M-1}\sum_{t=0}^{M-1-k} \mathbb{P}(\text{SINR}_{i,\mathbf{x}_\text{o}} > \beta \mid  n({\cal A}_{\mathbf{x}_\text{o}}) = k+t, n({\Phi}_{\mathbf{x}_\text{o}}^i) = k) \mathbb{P}(n({\Phi}_{\mathbf{x}_\text{o}}^i) = k)\mathbb{P}(n({{\Phi}_{\mathbf{x}_\text{o}} \backslash {\Phi}_{\mathbf{x}_\text{o}}^i})= t) \nonumber\\&+ \sum_{k=1}^{\infty}\sum_{t=M-k}^{\infty}\mathbb{P}(\text{SINR}_{i,\mathbf{x}_\text{o}} > \beta \mid n({\cal A}_{\mathbf{x}_\text{o}}) = M, n({\Phi}_{\mathbf{x}_\text{o}}^i) = k)\mathbb{P}(n({\Phi}_{\mathbf{x}_\text{o}}^i) = k)\mathbb{P}(n({\Phi}_{\mathbf{x}_\text{o}} \backslash {\Phi}_{\mathbf{x}_\text{o}}^i)= t) \nonumber\\
&\stackrel{(b)}{=} \sum_{k=1}^{M-1}\sum_{t=0}^{M-1-k} \mathbb{P}(\text{SINR}_{i,\mathbf{x}_\text{o}} > \beta \mid  n({\cal A}_{\mathbf{x}_\text{o}}) = k+t, n({\Phi}_{\mathbf{x}_\text{o}}^i) = k) \frac{e^{-\lambda \pi D^2}(\lambda \pi D^2)^{k+t}{p_i^{\mathbf{x}_\text{o}}}^k{(1-p_i^{\mathbf{x}_\text{o}})^t}}{k!t!}\nonumber\\&+ \sum_{k=1}^{\infty}\mathbb{P}(\text{SINR}_{i,\mathbf{x}_\text{o}} > \beta \mid n({\cal A}_{\mathbf{x}_\text{o}}) = M, n({\Phi}_{\mathbf{x}_\text{o}}^i) = k)\frac{e^{-\lambda p_i^{\mathbf{x}_\text{o}} \pi D^2}(\lambda p_i^{\mathbf{x}_\text{o}} \pi D^2)^k}{k!}\nonumber\\&\times\left(1-\sum_{t=0}^{M-1-k}\frac{e^{-\lambda (1-p_i^{\mathbf{x}_\text{o}}) \pi D^2}(\lambda (1-p_i^{\mathbf{x}_\text{o}}) \pi D^2)^t}{t!}\right),
\end{align}
where $(a)$ follows from $\mathbb{P}(\text{SINR}_{i,\mathbf{x}_\text{o}} > \beta \mid n({\Phi}_{\mathbf{x}_\text{o}}^i) = 0) = 0$, which denotes the case that there is no RU having $f_i^{\mathbf{x}_\text{o}}$, and the fact that there cannot be more than $M$ active RUs in $\Phi_{\mathbf{x}_\text{o}}$ according to the zero-forcing transmission strategy. Also, $(b)$ is obtained from the Poisson distributions $\mathbb{P}(n({\Phi}_{\mathbf{x}_\text{o}}^i) = k) = e^{-\lambda p_i^{\mathbf{x}_\text{o}} \pi D^2}\frac{(\lambda p_i^{\mathbf{x}_\text{o}} \pi D^2)^k}{k!}$ and $\mathbb{P}(n({\Phi}_{\mathbf{x}_\text{o}} \backslash {\Phi}_{\mathbf{x}_\text{o}}^i)= t)=e^{-\lambda (1-p_i^{\mathbf{x}_\text{o}}) \pi D^2}\frac{(\lambda (1-p_i^{\mathbf{x}_\text{o}}) \pi D^2)^t}{t!}$ and some straightforward simplifications. Then, to evaluate the probability terms in (12) for the general conditions $n({\cal A}_{\mathbf{x}_\text{o}}) = l$ and $n({\Phi}_{\mathbf{x}_\text{o}}^i) = k$, by conditioning on the serving distance $r$, we have
\begin{align}
 \mathbb{P}(\text{SINR}_{i,\mathbf{x}_\text{o}} > \beta \mid n({\cal A}_{\mathbf{x}_\text{o}}) = l, n({\Phi}_{\mathbf{x}_\text{o}}^i) = k) \nonumber\hspace{130pt}\\= \int_{0}^{D+\|\mathbf{x}_\text{o}\|} {\mathbb{P}}\left( {\frac{{P{g_z}{r^{ - \alpha_\text{i} }}}}{{{\sigma ^2} + {\cal I}_\text{o}}} > \beta } \mid n({\cal A}_{\mathbf{x}_\text{o}}) = l\right){f_{\|\mathbf{z}_{\text{CS}}\|}}\left( r\mid k\right)\mathrm{d}r,
\end{align}
where the inner integral probability can be expressed as
\begin{align}
{\mathbb{P}}\left( {\frac{{P{g_z}{r^{ - {\alpha_\text{i}} }}}}{{{\sigma ^2} + {\cal I}_\text{o}}} > \beta }\mid n({\cal A}_{\mathbf{x}_\text{o}}) = l \right) &= \mathbb{P}\left( {g_z > \frac{{\beta {r^{\alpha_\text{i}} }}}{P}\left( {{\sigma ^2} + {\cal I}_\text{o}} \right)}\mid n({\cal A}_{\mathbf{x}_\text{o}})=l \right)\nonumber\\ &\stackrel{(c)}{=} \mathbb{E}\left\{\mathop \sum \limits_{m = 0}^{M-l} \frac{{{{\left( {\frac{{\beta {r^{\alpha_\text{i}} }}}{P}\left( {{\cal I_\text{o}} + {\sigma ^2}} \right)} \right)}^m}}}{{m!}}\exp \left( - \frac{{\beta {r^{\alpha_\text{i}} }}}{P}\left( {{\cal I_\text{o}} + {\sigma ^2}} \right)\right)\right\}\nonumber\\
 &=\mathop \sum \limits_{m = 0}^{M-l} \frac{{{\beta ^m}{r^{{\alpha_\text{i}} m}}}}{{{P^m}m!}}\mathbb{E}\left\{ {{{\left( {{\cal I_\text{o}} + {\sigma ^2}} \right)}^m}{\exp\left( - \frac{{\beta {r^{\alpha_\text{i}} }}}{P}\left( {{\cal I_\text{o}} + {\sigma ^2}} \right)\right)}} \right\}\nonumber\\ &\stackrel{(d)}{=}\mathop \sum \limits_{m = 0}^{M-l} \frac{{{\beta ^m}{r^{{\alpha_\text{i}} m}}}}{{{P^m}m!}} (-1)^{m}\frac{\mathrm{d}^m e^{-s\sigma^2}{\cal L}_{\cal I_\text{o}}(s)}{\mathrm{d} s^m}\bigg|_{s = \frac{{\beta {r^{\alpha_\text{i}} }}}{P}},
\end{align}
where $(c)$ follows from the fact that $g_z$ is chi-squared with $2(M-l+1)$ degrees of freedom and $(d)$ follows from the definition of the LT of $\cal I_\text{o}$ as ${\cal L_{\cal I_\text{o}}}\left( s \right)=\mathbb{E}\left\{e^{-s {\cal I}_\text{o}}\right\}$ and the derivative property of the LT as $\mathbb{E}\left\{X^m e^{-sX}\right\} = (-1)^m \frac{\mathrm{d}^m {\cal L}_{X}(s)}{\mathrm{d} s^m}$ for a random variable $X$. 

In the following theorem, ${\cal L_{\cal I_\text{o}}}$ is characterized.

\begin{theorem}
The LT of the out of cloud interference is 
\begin{align}
{\cal L}_{\cal I_\text{o}}(s)=\exp\Biggl(-2\pi \lambda_\text{p}\int_{\max\left\{0,d_\text{g}-D\right\}}^{\max\left\{D,d_\text{g}-D\right\}}\biggl\{1-{\cal F}(s,x)\biggr\}x\mathrm{d}x-2\pi \lambda_\text{p}\int_{\max\left\{D,d_\text{g}-D\right\}}^{\infty}\biggl\{1-{\cal E}(s,x)\biggr\}x\mathrm{d}x\Biggr),
\end{align}
where
\begin{align}
{\cal F}(s,x) =\sum_{j=0}^{M-1}\left(\int_{d_\text{g}}^{\max\left\{D-x,d_\text{g}\right\}}\frac{\frac{2y}{D^2}}{{1 + sP{{{y}}^{ - {\alpha_\text{o}} }}}}\mathrm{d}y+\int_{\max\left\{D-x,d_\text{g}\right\}}^{D+x}\frac{\frac{{1}}{\pi D^2} \frac{{\partial {\cal B}_x(y) }}{{\partial y}}}{{1 + sP{{{y}}^{ - {\alpha_\text{o}} }}}}\mathrm{d}y\right)^j \frac{e^{-\lambda {\cal G}(x)}(\lambda {\cal G}(x))^j}{j!}\nonumber\\+\left(\int_{d_\text{g}}^{\max\left\{D-x,d_\text{g}\right\}}\frac{\frac{2y}{D^2}}{{1 + sP{{{y}}^{ - {\alpha_\text{o}} }}}}\mathrm{d}y+\int_{\max\left\{D-x,d_\text{g}\right\}}^{D+x}\frac{\frac{{1}}{\pi D^2} \frac{{\partial {\cal B}_x(y) }}{{\partial y}}}{{1 + sP{{{y}}^{ - {\alpha_\text{o}} }}}}\mathrm{d}y\right)^M \left(1-\sum_{j=0}^{M-1}\frac{e^{-\lambda {\cal G}(x)}(\lambda {\cal G}(x))^j}{j!}\right),
\end{align}
and
\begin{align}
{\cal E}(s,x) &=\sum_{j=0}^{M-1}\left(\int_{\max\left\{x-D,d_\text{g}\right\}}^{x+D}\frac{\frac{{1}}{\pi D^2} \frac{{\partial {\cal B}_x(y) }}{{\partial y}}}{{1 + sP{{{y}}^{ - {\alpha_\text{o}} }}}}\mathrm{d}y\right)^j \frac{e^{-\lambda {\cal G}(x)}(\lambda {\cal G}(x))^j}{j!}\nonumber\\&+\left(\int_{\max\left\{x-D,d_\text{g}\right\}}^{x+D}\frac{\frac{{1}}{\pi D^2} \frac{{\partial {\cal B}_x(y) }}{{\partial y}}}{{1 + sP{{{y}}^{ - {\alpha_\text{o}}}}}}\mathrm{d}y\right)^M \left(1-\sum_{j=0}^{M-1}\frac{e^{-\lambda {\cal G}(x)}(\lambda {\cal G}(x))^j}{j!}\right).
\end{align}
and
\begin{align}
{\cal G}(x) = \left\{ {\begin{array}{*{20}{c}}
{\pi D^2 - \pi (\min\left\{d_\text{g},D\right\})^2 \hspace{+15pt} 0\leq x<|D-d_\text{g}|,}\\
{\hspace{-8pt}\pi D^2 - {\cal B}_x(d_\text{g}) \hspace{+27pt} |D-d_\text{g}| \leq x<D+d_\text{g},}\\
{\hspace{-6pt}\pi D^2 \hspace{+131pt} x\ge D+d_\text{g},}
\end{array}} \right. 
\end{align}
where ${\cal B}_x$ is given in (10). 
\end{theorem}
\begin{IEEEproof}
See Appendix A.
\end{IEEEproof}

The derived conditional hit probability in (12) can be approximated with a simplified expression presented in the next lemma. 
\begin{lemma}
An approximation on the conditional hit probability given in (12) is
\begin{align}
\text{P}_{i,\mathbf{x}_\text{o}}^{\text{hit},\text{CS}} &\approx \mathop \sum \limits_{m = 1}^{{M-L+1}} {\left( { - 1} \right)^{m + 1}}
{{M-L+1}\choose{m}}\Biggl\{\int_{0}^{D-\|\mathbf{x}_\text{o}\|} 2\pi \lambda p_i^{\mathbf{x}_\text{o}} r e^{-\lambda p_i^{\mathbf{x}_\text{o}}\pi r^2} e^{ - \frac{\eta \beta m r^{{{\alpha_\text{i}}}}}{P} {\sigma ^2}}{\cal L}_{\cal I_\text{o}}\left(\frac{\eta\beta m r^{{{\alpha_\text{i}}}}}{P}\right) \mathrm{d}r\nonumber\\&+ \int_{D-\|\mathbf{x}_\text{o}\|}^{D+\|\mathbf{x}_\text{o}\|} \lambda p_i^{\mathbf{x}_\text{o}}\frac{{\partial {\cal B}_{\|\mathbf{x}_\text{o}\|}(r) }}{{\partial r}}{e^{ - \lambda p_i^{\mathbf{x}_\text{o}} {\cal B}_{\|\mathbf{x}_\text{o}\|}(r)}}e^{ - \frac{\eta \beta m r^{{{\alpha_\text{i}}}}}{P} {\sigma ^2}}{\cal L}_{\cal I_\text{o}}\left(\frac{\eta\beta m r^{{{\alpha_\text{i}}}}}{P}\right) \mathrm{d}r\Biggr\},
\end{align}
where $L = \min\left\{M,\left \lfloor{\lambda \pi D^2}\right \rfloor \right\}$ and $\eta ={ {{(M-L+1)}!}^{ - \frac{1}{{{(M-L+1)}}}}}$. Also, ${\cal L}_{\cal I_\text{o}}$ is given in (15).
\end{lemma}
\begin{IEEEproof}
According to the constraint $n({\cal A}_{\mathbf{x}_\text{o}}) = \min\left\{M,n(\Phi_{\mathbf{x}_\text{o}})\right\}$, we use this approximation that the number of active RUs in the representative cloud, i.e., $n({\cal A}_{\mathbf{x}_\text{o}})$, is closely equal to the average number of all the RUs in the cloud, i.e., $\lambda \pi D^2$, if $\lambda \pi D^2 < M$ and $n({\cal A}_{\mathbf{x}_\text{o}}) = M$ if $\lambda \pi D^2 \geq M$. Thus, we approximately have $n({\cal A}_{\mathbf{x}_\text{o}}) = \min\left\{M,\left \lfloor{\lambda \pi D^2}\right \rfloor\right\}$, where $\left \lfloor{\lambda \pi D^2}\right \rfloor$ is the closest integer smaller than or equal to $\lambda \pi D^2$. Then, defining $L = \min\left\{M,\left \lfloor{\lambda \pi D^2}\right \rfloor\right\}$ and from (12), we can write
\begin{align}
\text{P}_{i,\mathbf{x}_\text{o}}^{\text{hit},\text{CS}} \approx \sum_{k=1}^{\infty}\mathbb{P}(\text{SINR}_{i,\mathbf{x}_\text{o}} > \beta \mid  n({\cal A}_{\mathbf{x}_\text{o}}) = L, n({\Phi}_{\mathbf{x}_\text{o}}^i) = k)\mathbb{P}(n(\Phi_{\mathbf{x}_\text{o}}^i) = k)\nonumber\hspace{45pt}\\= \int_{0}^{D+\|\mathbf{x}_\text{o}\|} {\mathbb{P}}\left( {\frac{{P{g_z}{r^{ - {\alpha_\text{i}} }}}}{{{\sigma ^2} +\cal I_\text{o}}} > \beta } \mid n({\cal A}_{\mathbf{x}_\text{o}}) = L\right)\sum_{k=1}^{\infty}{f_{\|\mathbf{z}_{\text{CS}}\|}}\left( r\mid k\right)\mathbb{P}(n(\Phi_{\mathbf{x}_\text{o}}^i) = k)\mathrm{d}r.
\end{align}
The final result is obtained from 
\begin{align}
{\mathbb{P}}\left( {\frac{{P{g_z}{r^{ - {\alpha_\text{i}}}}}}{{{\sigma ^2} +\cal I_\text{o}}} > \beta } \mid n({\cal A}_{\mathbf{x}_\text{o}}) = L\right) \stackrel{(e)}{\approx} 1-\mathbb{E}\biggl\{\biggl(1-e^{-\frac{{\eta \beta {r^{\alpha_\text{i}} }}}{P}\left( {{\sigma ^2} + {\cal I}_\text{o}} \right)}\biggr)^{M-L+1}\biggr\}\nonumber\\\stackrel{(f)}{=} \mathop \sum \limits_{m = 1}^{{M-L+1}} {\left( { - 1} \right)^{m + 1}}
{{M-L+1}\choose{m}}e^{ - \frac{\eta \beta m r^{{{\alpha_\text{i}}}}}{P} {\sigma ^2}}{\cal L}_{\cal I_\text{o}}\left(\frac{\eta\beta m r^{{{\alpha_\text{i}}}}}{P}\right),
\end{align}
where $(e)$ is from the Alzer's lemma on chi-squared distributions with $2(M-L+1)$ degrees of freedom, where $\eta ={ {{(M-L+1)}!}^{ - \frac{1}{{{(M-L+1)}}}}}$ [40], and $(f)$ is from the binomial expansion. \\Also, the remaining term of the integrand in (20) can be computed as

\begin{align}
\sum_{k=1}^{\infty}{f_{\|\mathbf{z}_{\text{CS}}\|}}\left( r\mid k\right)\mathbb{P}(n(\Phi_{\mathbf{x}_\text{o}}^i) = k) = \sum_{k=1}^{\infty}kf^{\|\mathbf{x}_\text{o}\|}(r)(1-F^{\|\mathbf{x}_\text{o}\|}(r))^{k-1}\frac{e^{-\lambda p_i^{\mathbf{x}_\text{o}} \pi D^2}(\lambda p_i^{\mathbf{x}_\text{o}} \pi D^2)^k}{k!}\nonumber\hspace{10pt}\\
=(\lambda p_i^{\mathbf{x}_\text{o}} \pi D^2) f^{\|\mathbf{x}_\text{o}\|}(r)e^{-\lambda p_i^{\mathbf{x}_\text{o}} \pi D^2F^{\|\mathbf{x}_\text{o}\|}(r)} \sum_{k=1}^{\infty}e^{-\lambda p_i^{\mathbf{x}_\text{o}} \pi D^2(1-F^{\|\mathbf{x}_\text{o}\|}(r))}\frac{\bigl(\lambda p_i^{\mathbf{x}_\text{o}} \pi D^2(1-F^{\|\mathbf{x}_\text{o}\|}(r))\bigr)^{k-1}}{(k-1)!}  \nonumber
\end{align}
\vspace{-20pt}
\begin{align}
\stackrel{(g)}{=} \lambda p_i^{\mathbf{x}_\text{o}} \pi D^2 f^{\|\mathbf{x}_\text{o}\|}(r)e^{-\lambda p_i^{\mathbf{x}_\text{o}} \pi D^2F^{\|\mathbf{x}_\text{o}\|}(r)}\nonumber\hspace{136pt}\\\stackrel{(h)}{=} \left\{ {\begin{array}{*{20}{c}}
{{2\pi \lambda p_i^{\mathbf{x}_\text{o}} r e^{-\lambda p_i^{\mathbf{x}_\text{o}}\pi r^2}} \hspace{+99pt} 0\leq r<D-{\|\mathbf{x}_\text{o}\|},}\\
{{\lambda p_i^{\mathbf{x}_\text{o}}\frac{{\partial {\cal B}_{\|\mathbf{x}_\text{o}\|}(r) }}{{\partial r}}{e^{ - \lambda p_i^{\mathbf{x}_\text{o}} {\cal B}_{\|\mathbf{x}_\text{o}\|}(r)}}}\hspace{+15pt} D-{\|\mathbf{x}_\text{o}\|} \leq r<D+{\|\mathbf{x}_\text{o}\|},}\\
{0 \hspace{+198pt} r\ge D+{\|\mathbf{x}_\text{o}\|},}
\end{array}} \right. 
\end{align}
where $(g)$ is from the PDF of a Poisson random variable with mean $\lambda p_i^{\mathbf{x}_\text{o}} \pi D^2(1-F^{\|\mathbf{x}_\text{o}\|}(r))$ and $(h)$ is obtained from (8). 
\end{IEEEproof}

\subsection{Best Selection}

In the following theorem, the distribution of the power of the desired signal from the serving RU, i.e., $\mathop {\max }\limits_{{\mathbf{y}} \in \left\{{\Phi}_{\mathbf{x}_\text{o}}^i\mid n({\Phi}_{\mathbf{x}_\text{o}}^i)>0 \right\}} P g_y\| {{\mathbf{y}}} \|^{-{\alpha_\text{i}}}$, is characterized, which according to (5) and (7) is a key step in the sequel conditional hit probability analysis. 

\begin{theorem}
Given the conditions $n({\cal A}_{\mathbf{x}_\text{o}}) = l$ on the number of active RUs and $n({\Phi}_{\mathbf{x}_\text{o}}^i) = k$ on the number of RUs having $f_i^{\mathbf{x}_\text{o}}$, the CDF of the random variable $X = \mathop {\max }\limits_{{\mathbf{y}} \in \left\{{\Phi}_{\mathbf{x}_\text{o}}^i\mid n({\Phi}_{\mathbf{x}_\text{o}}^i)>0 \right\}} P g_y\| {{\mathbf{y}}} \|^{-{\alpha_\text{i}}}$ is given by
\begin{align}
F_X (x \mid n({\cal A}_{\mathbf{x}_\text{o}})=l, n({\Phi}_{\mathbf{x}_\text{o}}^i) = k) = \left(1-\mathop \sum \limits_{m = 0}^{M-l}\frac{x^m}{P^m m!}\int_{0}^{D+\|\mathbf{x}_\text{o}\|}{y} ^{{\alpha_\text{i}} m} {\rm{exp}}\left( { - \frac{x y^{{\alpha_\text{i}}}}{P}} \right)f^{\|\mathbf{x}_\text{o}\|}(y)\mathrm{d}y\right)^k,
\end{align}
where $f^{\|\mathbf{x}_\text{o}\|}$ is given in (8).
\end{theorem}
\begin{IEEEproof}
See Appendix B.
\end{IEEEproof}

Using Theorem 2 and similar to the approach in (12), we now derive the conditional hit probability of the user of interest for the best selection strategy as follows. 
\begin{align}
&\text{P}_{i,\mathbf{x}_\text{o}}^{\text{hit},\text{BS}}  = \mathbb{P}(\text{SINR}_{i,\mathbf{x}_\text{o}} > \beta)= 1-\mathbb{P}(\text{SINR}_{i,\mathbf{x}_\text{o}} < \beta) \nonumber\\
&=1-\sum_{k=0}^{M-1}\sum_{t=0}^{M-1-k} \mathbb{P}(\text{SINR}_{i,\mathbf{x}_\text{o}} < \beta \mid  n({\cal A}_{\mathbf{x}_\text{o}})=k+t, n({\Phi}_{\mathbf{x}_\text{o}}^i) = k) \mathbb{P}(n({\Phi}_{\mathbf{x}_\text{o}}^i) = k)\mathbb{P}(n({\Phi}_{\mathbf{x}_\text{o}}\backslash{\Phi}_{\mathbf{x}_\text{o}}^i)= t) \nonumber\\&- \sum_{k=0}^{\infty}\mathbb{P}(\text{SINR}_{i,\mathbf{x}_\text{o}} < \beta \mid n({\cal A}_{\mathbf{x}_\text{o}})=M, n({\Phi}_{\mathbf{x}_\text{o}}^i) = k)\mathbb{P}(n({\Phi}_{\mathbf{x}_\text{o}}^i) = k)\left(1-\sum_{t=0}^{M-1-k}\mathbb{P}(n({\Phi}_{\mathbf{x}_\text{o}}\backslash{\Phi}_{\mathbf{x}_\text{o}}^i) = t)\right),
\end{align}
where
\begin{align}
&\hspace{-50pt}\mathbb{P}(\text{SINR}_{i,\mathbf{x}_\text{o}} < \beta \mid n({\cal A}_{\mathbf{x}_\text{o}})=l, n({\Phi}_{\mathbf{x}_\text{o}}^i) = k) =\nonumber\\&\mathbb{P}\left(\mathop {\max }\limits_{{\mathbf{y}} \in \left\{{\Phi}_{\mathbf{x}_\text{o}}^i \mid n({\Phi}_{\mathbf{x}_\text{o}}^i) > 0\right\}} P g_y\| {{\mathbf{y}}} \|^{-{\alpha_\text{i}}}<\beta({\cal I}_\text{o}+\sigma^2)\mid n({\cal A}_{\mathbf{x}_\text{o}})=l, n({\Phi}_{\mathbf{x}_\text{o}}^i) = k\right) \nonumber
\end{align}
\begin{align}
&\stackrel{(i)}{=} \mathbb{E}\left\{\left(1-\mathop \sum \limits_{m = 0}^{M-l} \frac{\beta^m({\cal I}_\text{o}+\sigma^2)^m}{P^m m!}\int_{0}^{D+\|\mathbf{x}_\text{o}\|}{y} ^{{\alpha_\text{i}} m}{\rm{exp}}\left( { - \frac{\beta({\cal I}_\text{o}+\sigma^2) y^{{\alpha_\text{i}}}}{P}} \right)f^{\|\mathbf{x}_\text{o}\|}(y)\mathrm{d}y\right)^k \right\}\nonumber\\
 &\stackrel{(j)}{=} \int_{0}^{\infty}\left(1-\mathop \sum \limits_{m = 0}^{M-l} \frac{\beta^m(\gamma+\sigma^2)^m}{P^m m!}\int_{0}^{D+\|\mathbf{x}_\text{o}\|}{y} ^{{\alpha_\text{i}} m}{\rm{exp}}\left( { - \frac{\beta(\gamma+\sigma^2) y^{{\alpha_\text{i}}}}{P}} \right)f^{\|\mathbf{x}_\text{o}\|}(y)\mathrm{d}y\right)^k f_{{\cal I}_\text{o}}(\gamma) \mathrm{d}{\gamma},
\end{align}
where $(i)$ follows from the CDF of $\mathop {\max }\limits_{{\mathbf{y}} \in \left\{{\Phi}_{\mathbf{x}_\text{o}}^i \mid n({\Phi}_{\mathbf{x}_\text{o}}^i) >0\right\}} P g_y\| {{\mathbf{y}}} \|^{-{\alpha_\text{i}}}$ derived in Theorem 2 and the fact that $\mathbb{P}(\text{SINR}_{i,\mathbf{x}_\text{o}} < \beta \mid n({\Phi}_{\mathbf{x}_\text{o}}^i) = 0) = 1$ for $0 <\beta <\infty$, and in $(j)$ $f_{\cal I_\text{o}}$ is the PDF of the out of cloud interference $\cal I_\text{o}$. Using the Gil-Pelaez theorem [41], we can obtain $f_{\cal I_\text{o}}$ from its LT given in Theorem 1 as
\begin{eqnarray}
f_{\cal I_\text{o}} (\gamma) = \frac{1}{\pi} \int_{0}^{\infty}\Re \left\{e^{-j\gamma s} {\cal L}_{\cal I_\text{o}}(-js)\right\} \mathrm{d}s,
\end{eqnarray}
where $\Re\left\{z\right\}$ denotes the real part of the complex number $z$. However, it is computationally inefficient to evaluate (26). On the other hand, the Gamma distribution by matching first and second moments can provide an accurate approximation to the statistics of Poisson summations, as shown in [42]. Hence, as we have $\mathbb{E}\left\{\cal I_\text{o}\right\} = -\frac{\mathrm{d}{\cal L}_{\cal I_\text{o}}(0)}{\mathrm{d}s}$ and $\mathbb{E}\left\{{\cal I}_\text{o}^2\right\} = \frac{\mathrm{d}^2{\cal L}_{\cal I_\text{o}}(0)}{\mathrm{d}s^2}$, the approximated PDF of $\cal I_\text{o}$ can be computed as
\begin{eqnarray}
f_{\cal I_\text{o}} (\gamma) \approx \frac{{\gamma^{\tau-1} e^{-\frac{\gamma}{\zeta}}}}{{{\zeta}^\tau \Gamma (\tau)}},
\end{eqnarray}
where the shape parameter $\tau = \frac{{\mathbb{E}}^2\left\{{\cal I}_\text{o}\right\}}{\mathbb{E}\left\{{\cal I}_\text{o}^2\right\}-{\mathbb{E}}^2\left\{{\cal I}_\text{o}\right\}}$, the scale parameter $\zeta = \frac{\mathbb{E}\left\{{\cal I}_\text{o}^2\right\}-{\mathbb{E}}^2\left\{{\cal I}_\text{o}\right\}}{\mathbb{E}\left\{{\cal I}_\text{o}\right\}}$, and $\Gamma(\tau) = \int_{0}^{\infty} x^{\tau-1} e^{-x}\mathrm{d}x$ is the Gamma function.

Overally, the derived conditional hit probability in (24) can be approximated with a simplified expression presented in the next lemma.
\begin{lemma}
An approximation on the conditional hit probability given in (24) is
\begin{align}
\text{P}_{i,\mathbf{x}_\text{o}}^{\text{hit},\text{BS}}  \approx 1-\nonumber\hspace{400pt}\\\exp\Biggl(-\lambda p_i^{\mathbf{x}_\text{o}} \pi D^2\mathop \sum \limits_{m = 1}^{{M-L+1}} {\left( { - 1} \right)^{m + 1}}
{{M-L+1}\choose{m}}\int_{0}^{D+\|\mathbf{x}_\text{o}\|}e^{ - \frac{\eta \beta m y^{{{\alpha_\text{i}}}}}{P} {\sigma ^2}}{\cal L}_{\cal I_\text{o}}\left(\frac{\eta\beta m y^{{{\alpha_\text{i}}}}}{P}\right) f^{\|\mathbf{x}_\text{o}\|}(y)\mathrm{d}y\Biggr), 
\end{align}
where $L = \min\left\{M,\left \lfloor{\lambda \pi D^2}\right \rfloor \right\}$ and $\eta ={ {{(M-L+1)}!}^{ - \frac{1}{{{(M-L+1)}}}}}$. Also, $f^{\|\mathbf{x}_\text{o}\|}$ and $\cal L_{\cal I_\text{o}}$ are given in (8) and (15), respectively.
\end{lemma}
\begin{IEEEproof}
Similar to Lemma 2, we approximate $n({\cal A}_{\mathbf{x}_\text{o}})= L = \min\left\{M,\left \lfloor{\lambda \pi D^2}\right \rfloor \right\}$. Then, from (24) and (25), we have
\begin{align}
\text{P}_{i,\mathbf{x}_\text{o}}^{\text{hit},\text{BS}}  &\approx 1-\sum_{k=0}^{\infty}\mathbb{P}(n(\Phi_{\mathbf{x}_\text{o}}^i) = k)\mathbb{P}(\text{SINR}_{i,\mathbf{x}_\text{o}} < \beta \mid n({\cal A}_{\mathbf{x}_\text{o}})={L}, n({\Phi}_{\mathbf{x}_\text{o}}^i) = k) \nonumber\\ &= 1-\sum_{k=0}^{\infty}\mathbb{P}(n(\Phi_{\mathbf{x}_\text{o}}^i) = k)\mathbb{E}\Biggl\{\Biggl(1-\mathop \sum \limits_{m = 0}^{M-{L}} \frac{\beta^m({\cal I}_\text{o}+\sigma^2)^m}{P^m m!}\times\nonumber\\
&\int_{0}^{D+\|\mathbf{x}_\text{o}\|}{y} ^{{\alpha_\text{i}} m}{\rm{exp}}\left( { - \frac{\beta({\cal I}_\text{o}+\sigma^2) y^{{\alpha_\text{i}}}}{P}} \right)f^{\|\mathbf{x}_\text{o}\|}(y)\mathrm{d}y\Biggr)^k \Biggr\}= 1-\mathbb{E}\Biggl\{\sum_{k=0}^{\infty}\mathbb{P}(n(\Phi_{\mathbf{x}_\text{o}}^i) = k)\times\nonumber\\&\left(1-\mathop \sum \limits_{m = 0}^{M-{L}} \frac{\beta^m({\cal I}_\text{o}+\sigma^2)^m}{P^m m!}\int_{0}^{D+\|\mathbf{x}_\text{o}\|}{y} ^{{\alpha_\text{i}} m}{\rm{exp}}\left( { - \frac{\beta({\cal I}_\text{o}+\sigma^2) y^{{\alpha_\text{i}}}}{P}} \right)f^{\|\mathbf{x}_\text{o}\|}(y)\mathrm{d}y\right)^k \Biggr\}\nonumber\\
 &\stackrel{(k)}{=} 1-\mathbb{E}\Biggl\{\exp\Biggl(-\lambda p_i^{\mathbf{x}_\text{o}} \pi D^2\mathop \sum \limits_{m = 0}^{M-{L}} \frac{\beta^m({\cal I}_\text{o}+\sigma^2)^m}{P^m m!}\times\nonumber\\&\int_{0}^{D+\|\mathbf{x}_\text{o}\|}{y} ^{{\alpha_\text{i}} m}{\rm{exp}}\left( { - \frac{\beta({\cal I}_\text{o}+\sigma^2) y^{{\alpha_\text{i}}}}{P}} \right)f^{\|\mathbf{x}_\text{o}\|}(y)\mathrm{d}y\Biggr) \Biggr\} \stackrel{(l)}{\approx} 1-\exp\Biggl(-\lambda p_i^{\mathbf{x}_\text{o}} \pi D^2\times\nonumber\\&\int_{0}^{D+\|\mathbf{x}_\text{o}\|}\mathop \sum \limits_{m = 0}^{M-{L}} \frac{\beta^m y^{{\alpha_\text{i}} m}}{P^m m!}\mathbb{E}\left\{({\cal I}_\text{o}+\sigma^2)^m{\rm{exp}}\left( { - \frac{\beta({\cal I}_\text{o}+\sigma^2) y^{{\alpha_\text{i}}}}{P}} \right)\right\}f^{\|\mathbf{x}_\text{o}\|}(y)\mathrm{d}y\Biggr)\nonumber\\
&\stackrel{(m)}{=} 1-\exp\Biggl(-\lambda p_i^{\mathbf{x}_\text{o}} \pi D^2\int_{0}^{D+\|\mathbf{x}_\text{o}\|}\mathop \sum \limits_{m = 0}^{M-{L}} \frac{\beta^m y^{{\alpha_\text{i}} m}}{P^m m!}(-1)^m\frac{\mathrm{d}^m e^{-s\sigma^2}{\cal L}_{\cal I_\text{o}}(s)}{\mathrm{d} s^m}\bigg|_{s = \frac{{\beta {y^{\alpha_\text{i}} }}}{P}} f^{\|\mathbf{x}_\text{o}\|}(y)\mathrm{d}y\Biggr)\nonumber\\&\stackrel{(n)}{\approx} 1-\exp\Biggl(-\lambda p_i^{\mathbf{x}_\text{o}} \pi D^2\int_{0}^{D+\|\mathbf{x}_\text{o}\|}\mathop \sum \limits_{m = 1}^{{M-L+1}} {\left( { - 1} \right)^{m + 1}}
{{M-L+1}\choose{m}}e^{ - \frac{\eta \beta m y^{{{\alpha_\text{i}}}}}{P} {\sigma ^2}}\times\nonumber\\&\hspace{160pt}{\cal L}_{\cal I_\text{o}}\left(\frac{\eta\beta m y^{{{\alpha_\text{i}}}}}{P}\right) f^{\|\mathbf{x}_\text{o}\|}(y)\mathrm{d}y\Biggr),
\end{align}
where $(k)$ follows from the moment-generating function (MGF) of the number of points of $\Phi_{\mathbf{x}_\text{o}}^i$ with mean $\lambda p_i^{\mathbf{x}_\text{o}} \pi D^2$, $(l)$ is from approximating $\mathbb{E}\left\{\exp(-\epsilon(X))\right\}$ with its lower bound from the Jensen's inequality, i.e., $\exp(-\mathbb{E}\left\{\epsilon(X)\right\})$, for a random variable $X$ and a positive function $\epsilon$, $(m)$ follows from the derivative property of the LT, and $(n)$ uses the previously derived approximation (21) on the result in (14). 
\end{IEEEproof}
\begin{remark} \textit{The approximated results in (19) and (28) even in the case $\lambda \pi D^2 > M$, i.e., $L = M$, are dependent to the number of antennas $M$ since the LT of ${\cal I_\text{o}}$ in (15) is dependent to $M$.}
\end{remark}
\section{Content Caching Design}
In this section, we design the content caching placement for the representative cloud to maximize the hit probability, by replacing the conditional hit probability derived in (12) and (14) in (3), respectively for the closest and the best selections. However, since the equations (12) and (24) depend on the distance of the user of interest to the center of the representative cloud, and for a general design over the network, here we focus on the hit probability for a user located at the center of the cloud, i.e., $\|\mathbf{x}_\text{o}\|=0$. Furthermore, since the exact conditional hit probabilities are not tractable enough to be optimized and also not computationally efficient to be evaluated in each iteration of an optimization algorithm, instead we use the approximated ones given in (19) and (28) for the closest and the best selections, respectively. Then, using these approximations, in the following for each selection strategy, we find the cache probabilities $\cal P^{\mathbf{x}_\text{o}}$ that maximizes the hit probability assuming $\|\mathbf{x}_\text{o}\|=0$, through an optimization problem with the constraint on the memory size of caches. 

\textit{Problem 1:} For the closest selection, we have
\begin{align}
\max_{{\cal P}^{\mathbf{x}_\text{o}}} \sum_{i=1}^{N_{\mathbf{x}_\text{o}}} q_i^{\mathbf{x}_\text{o}}\mathop \sum \limits_{m = 1}^{{M-L+1}} {\left( { - 1} \right)^{m + 1}}{{M-L+1}\choose{m}}\int_{0}^{D} 2\pi \lambda p_i^{\mathbf{x}_\text{o}} r e^{-\lambda p_i^{\mathbf{x}_\text{o}}\pi r^2} e^{ - \frac{\eta \beta m r^{{{\alpha_\text{i}}}}}{P} {\sigma ^2}}{\cal L}_{\cal I_\text{o}}\left(\frac{\eta\beta m r^{{{\alpha_\text{i}}}}}{P}\right) \mathrm{d}r,
\end{align}
\hspace{10pt}subject to
\begin{align}
\sum\limits_{i = 1}^{N_{\mathbf{x}_\text{o}}} p_i^{\mathbf{x}_\text{o}} \le N_\text{c}^{\mathbf{x}_\text{o}},\hspace{310pt}\nonumber\\
0\leq p_i^{\mathbf{x}_\text{o}}\leq 1, \forall i=1,...,N_{\mathbf{x}_\text{o}}.\hspace{250pt}\nonumber
\end{align}

In the following theorem, we prove the concavity of Problem 1.
\begin{theorem}
Problem 1 is concave.
\end{theorem}
\begin{IEEEproof}
The constraints of Problem 1 are linear and then concave. See Appendix C that proves the concavity of the objective function.
\end{IEEEproof}

Since Problem 1 is concave, then we can efficiently find the optimal solution using the dual Lagrangian method [43]. The Lagrangian function is
\begin{align}
{\cal L}({\cal P}^{\mathbf{x}_\text{o}},\mu) = \sum_{i=1}^{N_{\mathbf{x}_\text{o}}} q_i^{\mathbf{x}_\text{o}}\mathop \sum \limits_{m = 1}^{{M-L+1}} {\left( { - 1} \right)^{m + 1}}{{M-L+1}\choose{m}}\int_{0}^{D} 2\pi \lambda p_i^{\mathbf{x}_\text{o}} r e^{-\lambda p_i^{\mathbf{x}_\text{o}}\pi r^2} e^{ - \frac{\eta \beta m r^{{{\alpha_\text{i}}}}}{P} {\sigma ^2}}{\cal L}_{\cal I_\text{o}}\left(\frac{\eta\beta m r^{{{\alpha_\text{i}}}}}{P}\right) \mathrm{d}r\nonumber\\-\mu \left(\sum\limits_{i = 1}^{N_{\mathbf{x}_\text{o}}} p_i^{\mathbf{x}_\text{o}} - N_\text{c}^{\mathbf{x}_\text{o}}\right).\hspace{250pt}
\end{align}
Then, through an iterative algorithm with step $\tau$, we update the cache probabilities ${\cal P}^{\mathbf{x}_\text{o}}$ and the Lagrangian factor $\mu$ at each iteration $t+1$, until its convergence, as
\begin{align}
\left\{\begin{matrix}
[\hat{\cal P}^{\mathbf{x}_\text{o}}, \mu]_{t+1} = [{\cal P}^{\mathbf{x}_\text{o}},\mu]_{t} +\tau \nabla{\cal L}({\cal P}^{\mathbf{x}_\text{o}},\mu),\\
[{\cal P}^{\mathbf{x}_\text{o}}]_{t+1} = [\hat{\cal P}^{\mathbf{x}_\text{o}}]_{t+1}\bigg|_0^1,\hspace{60pt}
\end{matrix}\right.
\end{align}
where $x \bigg|_0^1\stackrel{\Delta}{=} \left\{\begin{matrix}
0,\ \text{if} \ x< 0\hspace{20pt}\\
x,\ \text{if} \ 0\leq x\leq1\\
1,\ \text{if} \ x>1\hspace{20pt}
\end{matrix}\right.$ is to restrict the cache probabilities between 0 and 1, and
\begin{align}
\nabla{\cal L}({\cal P}^{\mathbf{x}_\text{o}},\mu) = \left[\frac{\partial {\cal L}({\cal P}^{\mathbf{x}_\text{o}},\mu)}{\partial p_1^{\mathbf{x}_\text{o}}},\cdots,\frac{\partial {\cal L}({\cal P}^{\mathbf{x}_\text{o}},\mu)}{\partial p_{N_{\mathbf{x}_\text{o}}}^{\mathbf{x}_\text{o}}},\frac{\partial {\cal L}({\cal P}^{\mathbf{x}_\text{o}},\mu)}{\partial \mu}\right],
\end{align}
where
\begin{align}
\hspace{-20pt}\left\{\begin{matrix}
\frac{\partial {\cal L}({\cal P}^{\mathbf{x}_\text{o}},\mu)}{\partial p_i^{\mathbf{x}_\text{o}}} = 2\pi \lambda q_i^{\mathbf{x}_\text{o}}\mathop \sum \limits_{m = 1}^{{M-L+1}} {\left( { - 1} \right)^{m + 1}}{{M-L+1}\choose{m}}\int_{0}^{D} (1-\lambda p_i^{\mathbf{x}_\text{o}}\pi r^2) r e^{-\lambda p_i^{\mathbf{x}_\text{o}}\pi r^2} e^{ - \frac{\eta \beta m r^{{{\alpha_\text{i}}}}}{P} {\sigma ^2}}{\cal L}_{\cal I_\text{o}}\left(\frac{\eta \beta m r^{{{\alpha_\text{i}}}}}{P}\right)\mathrm{d}r-\mu,\\ 
\frac{\partial {\cal L}({\cal P}^{\mathbf{x}_\text{o}},\mu)}{\partial \mu} = -\left(\sum\limits_{i = 1}^{N_{\mathbf{x}_\text{o}}} p_i^{\mathbf{x}_\text{o}} - N_\text{c}^{\mathbf{x}_\text{o}}\right).\hspace{339pt}
 \end{matrix}\right.
\end{align}
In (34), while the integrals can not be reduced to closed-form or further simplified, it is easy to evaluate them numerically as their ranges of integration are finite.

\textit{Problem 2:} For the best selection, we have
\begin{align}
\max_{{\cal P}^{\mathbf{x}_\text{o}}} \sum_{i=1}^{N_{\mathbf{x}_\text{o}}} q_i^{\mathbf{x}_\text{o}}\Biggl\{1 - \exp\Biggl(-\lambda p_i^{\mathbf{x}_\text{o}} \pi D^2 \mathop \sum \limits_{m = 1}^{{M-L+1}} \biggl\{{\left( { - 1} \right)^{m + 1}}
{{M-L+1}\choose{m}}\times\nonumber\\\int_{0}^{D}e^{ - \frac{\eta \beta m y^{{{\alpha_\text{i}}}}}{P} {\sigma ^2}}{\cal L}_{\cal I_\text{o}}\left(\frac{\eta\beta m y^{{{\alpha_\text{i}}}}}{P}\right) \frac{2y}{D^2}\mathrm{d}y\biggr\}\Biggr)\Biggr\},
\end{align}
\hspace{60pt}subject to
\begin{align}
\sum\limits_{i = 1}^{N_{\mathbf{x}_\text{o}}} p_i^{\mathbf{x}_\text{o}} \le N_\text{c}^{\mathbf{x}_\text{o}},\hspace{250pt}\nonumber\\
0\leq p_i^{\mathbf{x}_\text{o}}\leq 1, \forall i=1,...,N_{\mathbf{x}_\text{o}}.\hspace{185pt}\nonumber
\end{align}

The concavity of Problem 2 is proved in the following theorem.
\begin{theorem}
Problem 2 is a concave problem. 
\end{theorem}
\begin{IEEEproof}
We can re-write the following term in (35) as
\begin{align}
\mathop \sum \limits_{m = 1}^{{M-L+1}} {\left( { - 1} \right)^{m + 1}}
{{M-L+1}\choose{m}}\int_{0}^{D}e^{ - \frac{\eta \beta m y^{{{\alpha_\text{i}}}}}{P} {\sigma ^2}}{\cal L}_{\cal I_\text{o}}\left(\frac{\eta\beta m y^{{{\alpha_\text{i}}}}}{P}\right) \frac{2y}{D^2}\mathrm{d}y \nonumber\\ = \int_{0}^{D} \mathbb{E}\left\{\mathop \sum \limits_{m = 1}^{{M-L+1}} {\left( { - 1} \right)^{m + 1}}{{M-L+1}\choose{m}}e^{- \frac{\eta \beta m y^{{{\alpha_\text{i}}}}({\sigma ^2+{\cal I}_\text{o}})}{P}}\right\} \frac{2y}{D^2}\mathrm{d}y\nonumber\\
= \int_{0}^{D} \mathbb{E}\left\{1-\left(1-e^{- \frac{\eta \beta y^{{{\alpha_\text{i}}}}({\sigma ^2+{\cal I}_\text{o}})}{P}}\right)^{M-L+1}\right\} \frac{2y}{D^2}\mathrm{d}y,\hspace{64pt}
\end{align}
which is always positive since $M \geq L$ and $\left(1-e^{- \frac{\eta \beta y^{{{\alpha_\text{i}}}}({\sigma ^2+{\cal I}_\text{o}})}{P}}\right)^{M-L+1} \leq 1$. Hence, it is clear that the objective function of Problem 2 and its constraints are concave. 
\end{IEEEproof}

Here with no iterations, we can find the optimal solution using the Karush-Kuhn-Tucker
(KKT) conditions [43]. Consider the following Lagrangian function:
\begin{align}
{\cal L}({\cal P}^{\mathbf{x}_\text{o}},\mu) = \sum_{i=1}^{N_{\mathbf{x}_\text{o}}} q_i^{\mathbf{x}_\text{o}} \Biggl\{1 - \exp\Biggl(-\lambda p_i^{\mathbf{x}_\text{o}} \pi D^2 \mathop \sum \limits_{m = 1}^{{M-L+1}} \biggl\{{\left( { - 1} \right)^{m + 1}}
{{M-L+1}\choose{m}}\times\nonumber\\\int_{0}^{D}e^{ - \frac{\eta \beta m y^{{{\alpha_\text{i}}}}}{P} {\sigma ^2}}{\cal L}_{\cal I_\text{o}}\left(\frac{\eta\beta m y^{{{\alpha_\text{i}}}}}{P}\right) \frac{2y}{D^2}\mathrm{d}y\biggr\}\Biggr)\Biggr\}-\mu \left(\sum\limits_{i = 1}^{N_{\mathbf{x}_\text{o}}} p_i^{\mathbf{x}_\text{o}} - N_\text{c}^{\mathbf{x}_\text{o}}\right).
\end{align}
From $\frac{\partial {\cal L}}{\partial p_i^{\mathbf{x}_\text{o}}}=0$ and $0\leq p_i^{\mathbf{x}_\text{o}}\leq 1$, and some straightforward simplifications, the colsed-form optimal solution for $p_i^{\mathbf{x}_\text{o}}$ is obtained as
\begin{align}
{p_i^{\mathbf{x}_\text{o},*}} = \frac{v-\ln \left(\frac{1}{q_i^{\mathbf{x}_\text{o}}}\right)}{2\lambda \pi \mathop \sum \limits_{m = 1}^{{M-L+1}} {\left( { - 1} \right)^{m + 1}}
{{M-L+1}\choose{m}}\int_{0}^{D}ye^ { - \frac{\eta \beta m y^{{\alpha_\text{i}}}}{P}\sigma^2} {\cal L}_{\cal I_\text{o}}\left(\frac{\eta \beta m y^{{\alpha_\text{i}}}}{P}\right)\mathrm{d}y}\bigg|_0^1,
\end{align}
where $v$ can be obtained by the bisection method [43] from the constraint $\sum\limits_{i = 1}^{N_{\mathbf{x}_\text{o}}} p_i^{\mathbf{x}_\text{o},*} = N_\text{c}^{\mathbf{x}_\text{o}}$. 
 
\begin{corollary} In the best selection strategy, the content caching distribution tends to the uniform distribution, i.e., ${p_i^{\mathbf{x}_\text{o},*}} = \frac{N_\text{c}^{\mathbf{x}_\text{o}}}{N_{\mathbf{x}_\text{o}}}, \forall i$, when $\frac{2\lambda \pi \mathop \sum \limits_{m = 1}^{{M-L+1}} {\left( { - 1} \right)^{m + 1}}
{{M-L+1}\choose{m}}\int_{0}^{D}ye^ { - \frac{\eta \beta m y^{{\alpha_\text{i}}}}{P}\sigma^2} {\cal L}_{\cal I_\text{o}}\left(\frac{\eta \beta m y^{{\alpha_\text{i}}}}{P}\right)\mathrm{d}y}{\ln \left(\frac{q_1^{\mathbf{x}_\text{o}}}{q_{{N_{\mathbf{x}_\text{o}}}}^{\mathbf{x}_\text{o}}}\right)}$ tends to $\infty$.
\end{corollary}
\begin{IEEEproof} According to (38), the gap between each two cache probabilities $p_i^{\mathbf{x}_\text{o},*}$ and $p_j^{\mathbf{x}_\text{o},*}$ is \\$\frac{\ln \left(\frac{q_i^{\mathbf{x}_\text{o}}}{q_j^{\mathbf{x}_\text{o}}}\right)}{2\lambda \pi \mathop \sum \limits_{m = 1}^{{M-L+1}} {\left( { - 1} \right)^{m + 1}}
{{M-L+1}\choose{m}}\int_{0}^{D}ye^ { - \frac{\eta \beta m y^{{\alpha_\text{i}}}}{P}\sigma^2} {\cal L}_{\cal I_\text{o}}\left(\frac{\eta \beta m y^{{\alpha_\text{i}}}}{P}\right)\mathrm{d}y}$. Hence, all cache probabilities tend to be equal when the maximum gap, which is between $p_1^{\mathbf{x}_\text{o},*}$ and $p_{N_{\mathbf{x}_\text{o}}}^{\mathbf{x}_\text{o},*}$, tends to 0.
\end{IEEEproof}
\begin{remark} \textit{Considering the special case of uniform popularity distribution, i.e., $q_i^{\mathbf{x}_\text{o}} = \frac{1}{N_{\mathbf{x}_\text{o}}}, \forall i$, 
it is clear from (34), (38), and the memory size constraint that the optimal content caching distribution is uniform, i.e., ${p_i^{\mathbf{x}_\text{o},*}} = \frac{N_\text{c}^{\mathbf{x}_\text{o}}}{N_{\mathbf{x}_\text{o}}}, \forall i$, for both selection strategies.}
\end{remark}
\section{Results and Discussion}
In this section, we provide numerical and simulation results for a specific scenario of wireless cloud caching networks with the setting parameters given in TABLE I, unless otherwise stated. For the notation simplicity, here we ignore the index $\mathbf{x}_\text{o}$ for the related parameters to the representative cloud. Also, we define $d = \|\mathbf{x}_\text{o}\|$.
\begin{table}[t]
\caption {Parameter Values} 
\vspace{-10pt}
\begin{center}
\resizebox{4.4cm}{!} {
    \begin{tabular}{| l | l |}
  
   \hline
    \hline
    \textbf{Setting Parameter} &{\textbf{Value}} \\ \hline
    $P$ & 0 dB\\ \hline
    ${\alpha_\text{i}}$ & 2.5\\ \hline
    ${\alpha_\text{o}}$ & 3\\ \hline
    $\beta$ & 10 dB\\ \hline
    $\lambda_{\rm p}$ & 0.0001 $\text{m}^{-2}$\\ \hline
    $\lambda$ & 0.1 $\text{m}^{-2}$\\ \hline
    $\sigma^2$ & -30 dB \\ \hline
    $\gamma$ & 0.7 \\ \hline
    $D$ & 30 m \\ \hline
    $d_\text{g}$ & 2 m \\ \hline
    $d$ & 10 m \\ \hline
    $N$ & 20\\ \hline
    $N_\text{c}$ & 10 \\ \hline
    $M$ & 10 \\ \hline
   \hline
    \end{tabular}}
 
\end{center}
\vspace{-15pt}
\end{table}

We provide Monte Carlo simulations to validate the accuracy of the exact and approximate results. In addition, we discuss the results and provide key design insights, based on the impacts of the distance of the user of interest from the center of the representative cloud, the memory size, the skewness parameter of the Zipf popularity distribution, the number of antennas, and the pathloss exponent on the hit probability. The benchmark caching strategy versus our "probabilistic" content placement is the "most popular" content placement that caches the most popular files as a greedy scheme. 

\begin{figure}[tb!]
\centering
\includegraphics[width =5.0in]{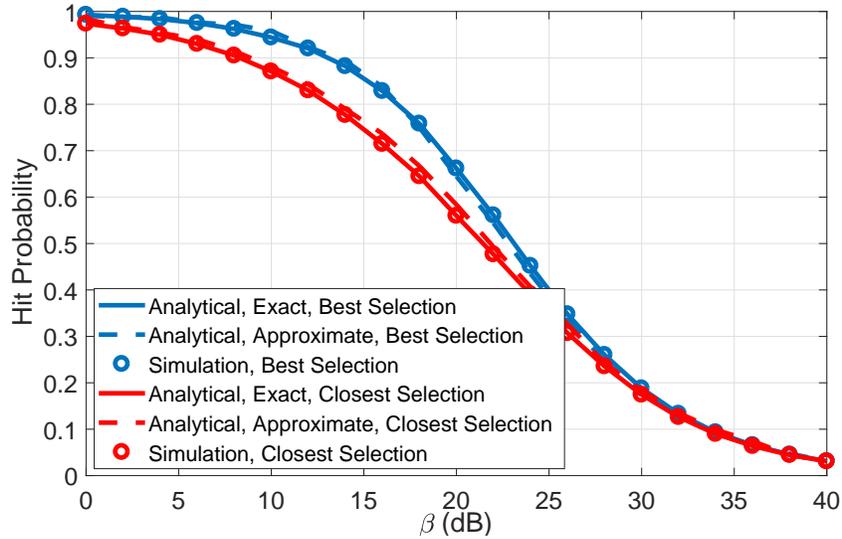} 
\caption{Hit probability as a function of the SINR threshold $\beta$.}
\end{figure}

\begin{figure}[t!]
\centering
\includegraphics[width =5.0in]{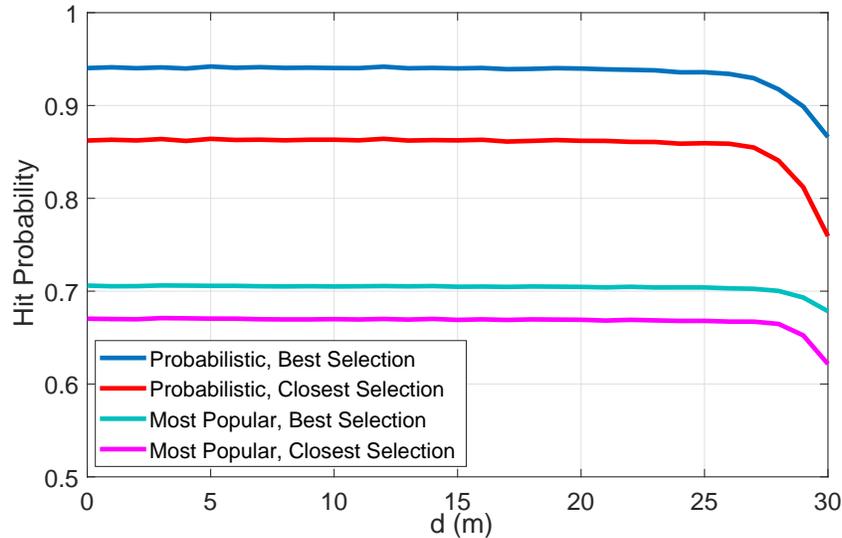}
\caption{Hit probability as a function of the distance $d$ of the user from the center of the representative cloud.}
\end{figure}

\begin{figure}[t!]
\centering
\includegraphics[width =5.0in]{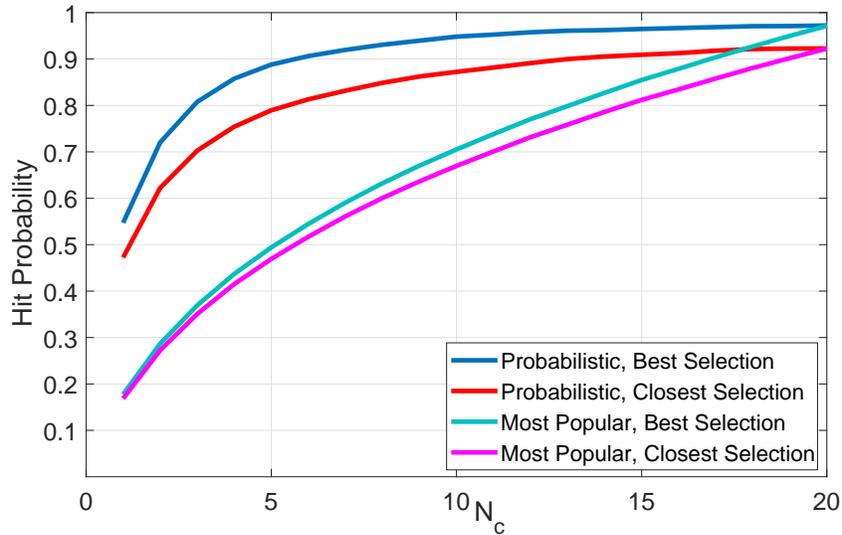}
\caption{Hit probability as a function of the memory size $N_\text{c}$.}
\end{figure}

\begin{figure}[t!]
\centering
\includegraphics[width =5.0in]{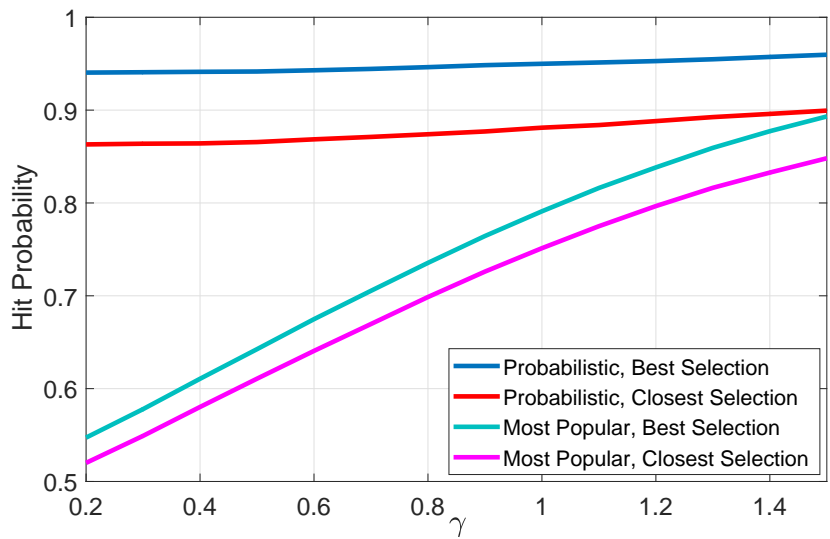}
\caption{Hit probability as a function of the Zipf skewness parameter $\gamma$.}
\end{figure}

\begin{figure}[t!]
\centering
\includegraphics[width =5.0in]{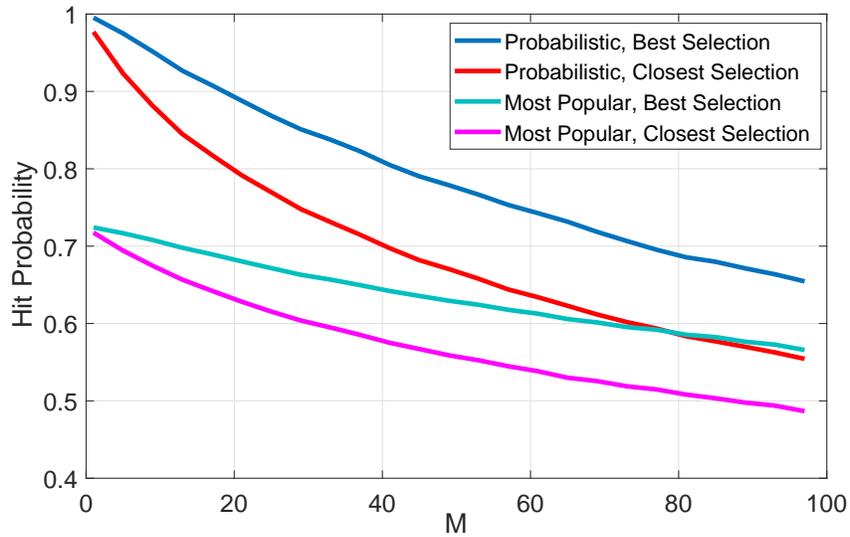}
\caption{Hit probability as a function of the number of antennas $M$.}
\end{figure}

\begin{figure}[t!]
\centering
\includegraphics[width =5.0in]{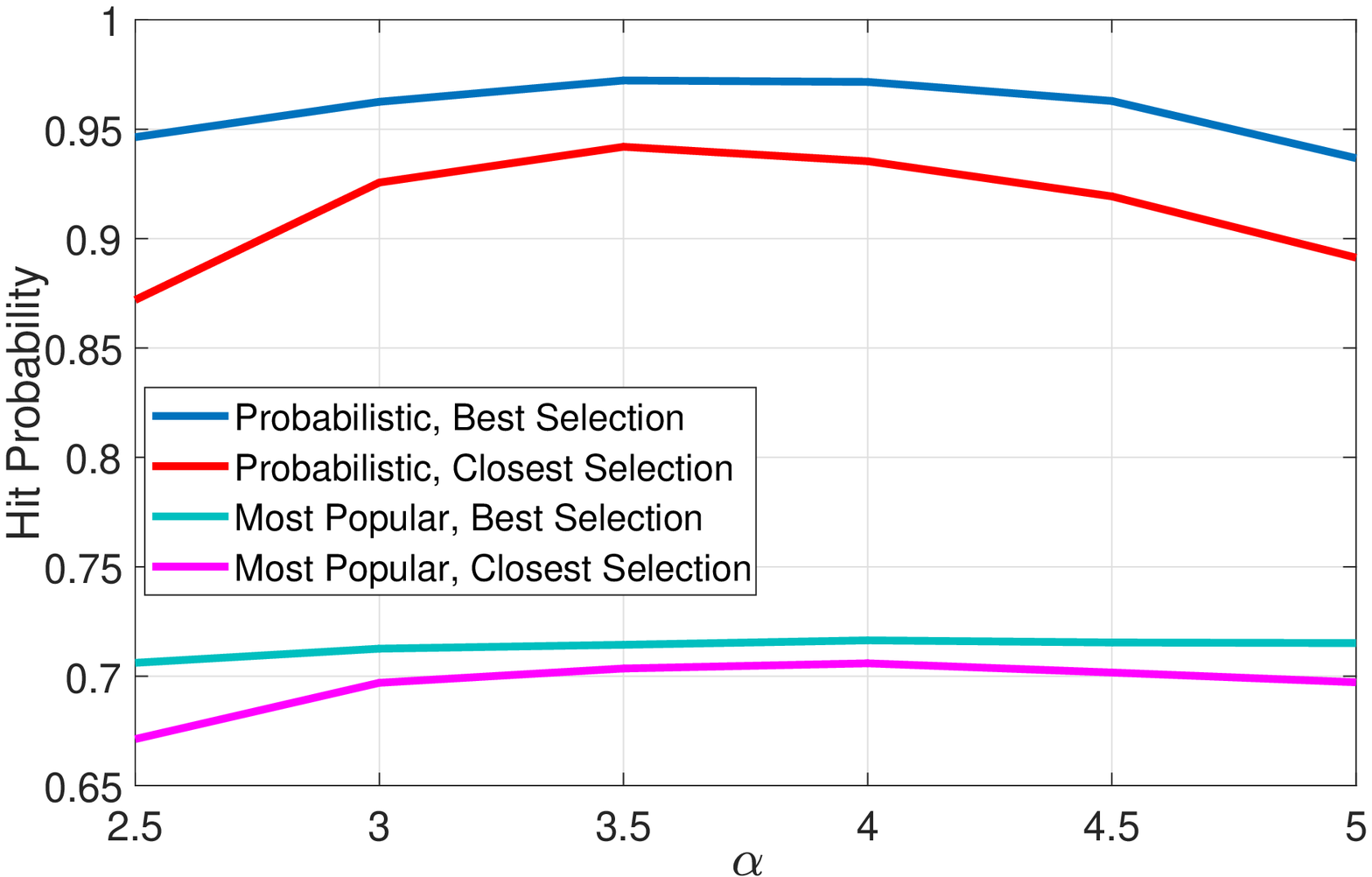}
\caption{Hit probability as a function of the pathloss exponent ${\alpha_\text{i}} = \alpha$.}
\end{figure}

\textit{Validation:} In Fig. 2, the analytical exact and approximate results and Monte Carlo simulations for the hit probability are shown as a function of the minimum required threshold $\beta$ for the closest and best selection strategies. It is observed that the analytical exact and approximate results precisely and tightly mimic the Monte Carlo simulations, respectively. Also, the best selection outperforms the closest selection at all $\beta$. For example, there is around 4 dB gap between the selection strategies at a target hit probability 0.8. 

\textit{Effect of the user distance from the center of the representative cloud:} The hit probability as a function of the distance $d$ is studied in Fig. 3 for both selection strategies. It is observed that the hit probability is maximized and minimized at the center and the boundary of the representative cloud, respectively. This is because our caching design is to maximize the hit probability at the center of the cloud. Also, as a degrading effect on the hit probability, the desired received powers of the RUs selected from both selection strategies decrease as the distance of the user from the center of the cloud increases. This observation and the low difference between the maximum
and minimum of the hit probability justify that the assumption of the user of interest being at the center of the cloud can well be applied for the content caching design. We further observe that our probabilistic approach significantly outperforms the most popular approach. This is due to the fact that the most popular approach is a specefic case of the probabilistic content placement when the cache probabilities of the $N_\text{c}$ most popular files are 1 and the rest are zero, where $N_\text{c}$ is the memory size.

\textit{Effect of the memory size:} The hit probability as a function of the memory size $N_\text{c}$ is shown in Fig. 4. As observed, the hit probability increases and the most popular approach proceeds towards the probabilistic approach as the meomery size increases, which is aligned with intuition. Also, from this figure, the probabilistic approach requires 10 less memory units (out of the total library size 20) than the most popular approach to achieve the same target hit probability 0.8. 
 
\textit{Effect of the Zipf skewness parameter:} In Fig. 5, the hit probability is plotted as a function of the skewness parameter $\gamma$. We observe that the hit probability increases with $\gamma$. This is because the probabilistic content placement as a popularity-aware design desires to cache more popular files in the network and the probability that a requested file is a popular one increases when $\gamma$ increases. Also, we observe that the most popular content placement is more dependent on $\gamma$. This is due to the fact that the not stored files over the caches are requested with lower probabilities as $\gamma$ increases. 

\textit{Effect of the number of antennas:} The hit probability as a function of the number of antennas $M$ is evaluated in Fig. 6. It is observed that the hit probability is degraded by increasing the number of antennas. This is due to the fact that having more antennas means considering less resource channels in the network and hence more interference from other clouds as the number of interfering RUs of each cloud increases with the number of antennas. 

\textit{Effect of the pathloss exponent:} In Fig. 7, the hit probability as a function of the pathloss exponent $\alpha_\text{i} = {\alpha}$ is plotted, considering $\alpha_\text{o} = {\alpha}+0.5$. An optimal value for ${\alpha_\text{i}}$ around 3.5 is observed in terms of the hit probability for both selection strategies. This is due to the fact that the SINR has a tradeoff since the power of both the desired and the interfering signals decrease as the pathloss exponent increases.

\section{Conclusion}
In this paper, using the Matern cluster process and the probabilistic content placement, we modeled wireless cloud caching networks comprised of distributedly coordinated multiple-antenna RUs. Suitable for different wireless applications, we considered two strategies for a user of interest to select a RU that has its desirable file over a representative cloud: closest selection and best selection. Accordingly, we derived exact expressions for the hit probability. We also approximated the hit probabilities with more convenient ones for caching optimizations. Then, for both selection strategies, we provided efficient algorithms to find the content placements maximizing the hit probabilities under the memory size constraint. 

Our analysis revealed that decreasing the number of antennas or the distance of the user from the center of its cloud and also increasing the skewness of the popularity distribution or the memory size improve the performance in terms of the hit probability. In addition, there exists a pathloss exponent for channels that leads to the highest performance. Also, the probabilistic content placement significantly outperforms the most popular content placement where only most popular files are stored. Moreover, the best selection strategy shows better performance than the closest selection strategy.

\appendices

\section{Proof of Theorem 1}
The LT of the out of cloud interference is
\begin{align}
&{\cal L}_{{\cal I}_\text{o}} (s) =\mathbb{E}\Biggl\{\exp\Biggl(-s \mathop \sum \limits_{\mathbf{x}\in {\Phi}_{\rm p}}\mathop \sum \limits_{{{\mathbf{y}} \in {{\cal A} _\mathbf{x}}}} P{f_{{{y}}}}{\| {{\mathbf{y}}} \|^{ - {\alpha_\text{o}} }}\Biggr)\Biggr\}=\mathbb{E}\Biggl\{\mathop \prod \limits_{{\mathbf{x}\in {\Phi}_{\rm p}}}\mathop \prod \limits_{\mathbf{y} \in {\cal A}_\mathbf{x} } \exp\left(-s P{f_{{{y}}}}{\| {{\mathbf{y}}} \|^{ - {\alpha_\text{o}} }}\right)\Biggr\}\nonumber\hspace{0pt}\\
&\stackrel{(a)}{=}\mathbb{E}\Biggl\{\mathop \prod \limits_{{\mathbf{x}\in {\Phi}_{\rm p}}}\mathop \prod \limits_{\mathbf{y} \in {\cal A}_\mathbf{x} }\frac{1}{{1 + sP{{\| \mathbf{y} \|}^{ - {\alpha_\text{o}} }}}}\Biggr\} \stackrel{(b)}{=} \mathbb{E}\Biggl\{\mathop \prod \limits_{{\mathbf{x}\in {\Phi}_{\rm p}}}\left(\int_{d_\text{g}}^{D+\|\mathbf{x}\|}\frac{1}{{1 + sP{{{y}}^{ - {\alpha_\text{o}} }}}}f^{\|\mathbf{x}\|}(y)\mathrm{d}y\right)^{n({\cal A}_\mathbf{x})}\mid n({\cal A}_\mathbf{x})\Biggr\}\nonumber\\
 &\stackrel{(c)}{=} \mathbb{E}\Biggl\{\mathop \prod \limits_{{\mathbf{x}\in {\Phi}_{\rm p}}}\sum_{j=0}^{M-1}\left(\int_{d_\text{g}}^{D+\|\mathbf{x}\|}\frac{1}{{1 + sP{{{y}}^{ - {\alpha_\text{o}} }}}}f^{\|\mathbf{x}\|}(y)\mathrm{d}y\right)^j \mathbb{P}(n({\cal A}_\mathbf{x}) = j)\nonumber\\&+\left(\int_{d_\text{g}}^{D+\|\mathbf{x}\|}\frac{1}{{1 + sP{{{y}}^{ - {\alpha_\text{o}} }}}}f^{\|\mathbf{x}\|}(y)\mathrm{d}y\right)^M \left(1-\sum_{j=0}^{M-1}\mathbb{P}(n({\cal A}_\mathbf{x}) = j)\right)\Biggr\} \stackrel{(d)}{=} \exp\Biggl(-\lambda_\text{p}\int_{\mathbb{R}^2\backslash \mathbf{b}(\mathbf{o},\max\left\{0,d_\text{g}-D\right\})}^{}\nonumber\\&\biggl\{1-\sum_{j=0}^{M-1}\left(\int_{d_\text{g}}^{D+\|\mathbf{x}\|}\frac{1}{{1 + sP{{{y}}^{ - {\alpha_\text{o}} }}}}f^{\|\mathbf{x}\|}(y)\mathrm{d}y\right)^j \frac{e^{-\lambda |\mathbf{b}(\mathbf{x},D)\backslash \mathbf{b}(\mathbf{o},d_\text{g})|}\left(\lambda |\mathbf{b}(\mathbf{x},D)\backslash \mathbf{b}(\mathbf{o},d_\text{g})|\right)^j}{j!}-\nonumber\\&\left(\int_{d_\text{g}}^{D+\|\mathbf{x}\|}\frac{1}{{1 + sP{{{y}}^{ - {\alpha_\text{o}} }}}}f^{\|\mathbf{x}\|}(y)\mathrm{d}y\right)^M \left(1-\sum_{j=0}^{M-1}\frac{e^{-\lambda |\mathbf{b}(\mathbf{x},D)\backslash \mathbf{b}(\mathbf{o},d_\text{g})|}\left(\lambda |\mathbf{b}(\mathbf{x},D)\backslash \mathbf{b}(\mathbf{o},d_\text{g})|\right)^j}{j!}\right)\biggr\}\mathrm{d}\mathbf{x}\Biggr),
\end{align}
where $(a)$ follows from $f_y \sim \exp(1)$, $(b)$ is due the facts that ${\cal A}_{\mathbf{x}}$ conditioned on $n({\cal A}_{\mathbf{x}})$ is a BPP where the distance of each point is i.i.d. with distribution $f^{\|{\mathbf{x}}\|}$ and no point can have distance less than $d_\text{g}$ to the user of interest, and $(c)$ is because there cannot be more than $M$ active RUs in ${\cal A}_\mathbf{x}$. Also, $(d)$ comes from the PGFL of the PPP [12, Thm. 4.9], the fact that clouds with $\|\mathbf{x}\|\leq \max\left\{0,d_\text{g}-D\right\}$ are completely inside the guard disk $\mathbf{b}(\mathbf{o},d_\text{g})$ and have no active RUs, and the Poisson distribution $\mathbb{P}(n({\cal A}_\mathbf{x}) = j) = \frac{e^{-\lambda |\mathbf{b}(\mathbf{x},D)\backslash \mathbf{b}(\mathbf{o},d_\text{g})|}\left(\lambda |\mathbf{b}(\mathbf{x},D)\backslash \mathbf{b}(\mathbf{o},d_\text{g})|\right)^j}{j!}$ where $\mathbf{b}(\mathbf{x},D)\backslash \mathbf{b}(\mathbf{o},d_\text{g})$ denotes the region that the interfering active RUs of the cloud ${\Phi}_{\mathbf{x}}$ are distributed over. We have that the area $|\mathbf{b}(\mathbf{x},D)\backslash \mathbf{b}(\mathbf{o},d_\text{g})| = \pi D^2-|\mathbf{b}(\mathbf{x},D)\cap \mathbf{b}(\mathbf{o},d_\text{g})|$, where the intersection area $|\mathbf{b}(\mathbf{x},D)\cap \mathbf{b}(\mathbf{o},d_\text{g})|$ for $|D-d_\text{g}| \leq \|\mathbf{x}\|<D+d_\text{g}$ can be computed by ${\cal B}_{\|\mathbf{x}\|}(d_\text{g})$ given in (10) and for other values of $\|\mathbf{x}\|$ it can be one of $\left\{0,\pi d_\text{g}^2,\pi D^2\right\}$. Finally, the presented result in (15) is obtained by converting (39) from Cartesian
to polar coordinates, replacing $f^{\|\mathbf{x}\|}$ with its given results in (8) and (9), and some straightforward simplifications.

\section{Proof of Theorem 2}
We can derive the CDF as follows.
\begin{align}
&\mathbb{P}\left(\mathop {\max }\limits_{{\mathbf{y}} \in \left\{{\Phi}_{\mathbf{x}_\text{o}}^i\mid n({\Phi}_{\mathbf{x}_\text{o}}^i)>0 \right\}} P g_y\| {{\mathbf{y}}} \|^{-{\alpha_\text{i}}} < x\mid n({\cal A}_{\mathbf{x}_\text{o}})=l, n({\Phi}_{\mathbf{x}_\text{o}}^i) = k\right)\nonumber\\ &\stackrel{(a)}{=} \mathbb{E}\left\{\mathop \prod \limits_{{\mathbf{y}} \in {\Phi}_{\mathbf{x}_\text{o}}^i} \mathbb{P}\left( P g_y\| {{\mathbf{y}}} \|^{-{\alpha_\text{i}}} < x \mid n({\cal A}_{\mathbf{x}_\text{o}})=l\right)\mid n({\Phi}_{\mathbf{x}_\text{o}}^i) = k\right\} \nonumber\\&=  \mathbb{E}\left\{\mathop \prod \limits_{{\mathbf{y}} \in {\Phi}_{\mathbf{x}_\text{o}}^i } \mathbb{P}\left( g_y <  \frac{x\| {{\mathbf{y}}} \|^{{\alpha_\text{i}}}}{P}\mid n({\cal A}_{\mathbf{x}_\text{o}})=l\right)\mid n({\Phi}_{\mathbf{x}_\text{o}}^i) = k\right\}\nonumber\\&\stackrel{(b)}{=} \mathbb{E}\left\{\mathop \prod \limits_{{\mathbf{y}} \in {\Phi}_{\mathbf{x}_\text{o}}^i}\left(1-\mathop \sum \limits_{m = 0}^{M-l} \frac{{{{\left( {\frac{x \| {{\mathbf{y}}} \|^{{\alpha_\text{i}}}}{P}} \right)}^m}}}{{m!}}{\rm{exp}}\left( { - \frac{x \| {{\mathbf{y}}} \|^{{\alpha_\text{i}}}}{P}} \right)\right)\mid n({\Phi}_{\mathbf{x}_\text{o}}^i) = k\right\}\nonumber\\
 &\stackrel{(c)}{=} \left(\int_{0}^{D+\|\mathbf{x}_\text{o}\|} \left\{1-\mathop \sum \limits_{m = 0}^{M-l} \frac{x^m {y} ^{{\alpha_\text{i}} m}}{P^m m!}{\rm{exp}}\left( { - \frac{x y^{{\alpha_\text{i}}}}{P}} \right)\right\} f^{\|\mathbf{x}_\text{o}\|}(y)\mathrm{d}y\right)^k,
\end{align} 
where $(a)$ follows from the independency of random variables $\left\{g_y\right\}$, $(b)$ follows from the fact that $g_y$ is chi-squared with $2(M-l-1)$ degrees of freedom, and $(c)$ is from the fact that ${\Phi}_{\mathbf{x}_\text{o}}^i$ conditioned on $n({\Phi}_{\mathbf{x}_\text{o}}^i)$ is a BPP where each point has distance distribution $f^{\|{\mathbf{x}_\text{o}}\|}$ in (8).

\section{Proof of Theorem 3}
We prove that the objective function of (30) for each $i$ is concave, and then we can conclude that the entire objective function is concave. 

We can re-write the objective function for each $i$ as
\begin{align}
\mathop \sum \limits_{m = 1}^{{M-L+1}} {\left( { - 1} \right)^{m + 1}}{{M-L+1}\choose{m}}\int_{0}^{D} 2\pi \lambda p_i^{\mathbf{x}_\text{o}} r e^{-\lambda p_i^{\mathbf{x}_\text{o}}\pi r^2} e^{ - \frac{\eta \beta m r^{{{\alpha_\text{i}}}}}{P} {\sigma ^2}}{\cal L}_{\cal I_\text{o}}\left(\frac{\eta\beta m r^{{{\alpha_\text{i}}}}}{P}\right) \mathrm{d}r =\nonumber\hspace{20pt}\\ \int_{0}^{\infty}\left\{\int_{0}^{D} 2\pi \lambda p_i^{\mathbf{x}_\text{o}} r e^{-\lambda p_i^{\mathbf{x}_\text{o}}\pi r^2} \mathop \sum \limits_{m = 1}^{{M-L+1}} {\left( { - 1} \right)^{m + 1}}{{M-L+1}\choose{m}} e^{ - \frac{\eta \beta m r^{{{\alpha_\text{i}}}}}{P} {(\sigma ^2+\gamma)}}\mathrm{d}r\right\} f_{\cal I_\text{o}}(\gamma)\mathrm{d}\gamma \nonumber\\= \int_{0}^{\infty}\left\{\int_{0}^{D} 2\pi \lambda p_i^{\mathbf{x}_\text{o}} r e^{-\lambda p_i^{\mathbf{x}_\text{o}}\pi r^2} \Biggl(1-\biggl(1- e^{ - \frac{\eta \beta  r^{{{\alpha_\text{i}}}}}{P} {(\sigma ^2+\gamma)}}\biggr)^{M-L+1}\Biggr)\mathrm{d}r\right\} f_{\cal I_\text{o}}(\gamma)\mathrm{d}\gamma,
\end{align}
where $f_{\cal I_\text{o}}$ is the PDF of ${\cal I}_\text{o}$ and the final result is obtained from the binomial expansion. Then, if we prove that the function ${\cal K}_i (\gamma)= \int_{0}^{D} 2\pi \lambda p_i^{\mathbf{x}_\text{o}} r e^{-\lambda p_i^{\mathbf{x}_\text{o}}\pi r^2} \bigl(1-\bigl(1- e^{ - \frac{\eta \beta r^{{{\alpha_\text{i}}}}}{P} {(\sigma ^2+\gamma)}}\bigr)^{M-L+1}\bigr)\mathrm{d}r$ for any positive $\gamma$ is concave, then we can prove that (41) which is a linear combination of ${\cal K}_i(\gamma)$ with the positive coefficient $f_{\cal I_\text{o}}(\gamma)$ for different positive values of $\gamma$ is concave.

To prove the concavity of ${\cal K}_i(\gamma)$, it is enough to show that the second order derivation of ${\cal K}_i(\gamma)$ is negative. By straightforward computations, this claim is given by
\begin{align}
\frac{\partial {\cal K}_i^2(\gamma)}{\partial^2 p_i^{\mathbf{x}_\text{o}}} = -\int_{0}^{D} 2\pi^2 \lambda^2 r^3 (2-\lambda \pi r^2 p_i^{\mathbf{x}_\text{o}}) e^{-\lambda p_i^{\mathbf{x}_\text{o}} \pi r^2} \biggl(1-\biggl(1- e^{ - \frac{\eta \beta  r^{{{\alpha_\text{i}}}}}{P} {(\sigma ^2+\gamma)}}\biggr)^{M-L+1}\biggr) \mathrm{d}r \leq 0,
\end{align}
which, by replacing the variable $x = r^2$, leads to the following inequality
\begin{align}
\int_{0}^{D^2} x e^{-\lambda p_i^{\mathbf{x}_\text{o}} \pi x} \biggl(1-\biggl(1- e^{ - \frac{\eta \beta  x^{\frac{{\alpha_\text{i}}}{2}}}{P} {(\sigma ^2+\gamma)}}\biggr)^{M-L+1}\biggr)\mathrm{d}x \geq \nonumber\hspace{100pt}\\ \frac{\lambda \pi p_i^{\mathbf{x}_\text{o}}}{2}\int_{0}^{D^2} x^2 e^{-\lambda p_i^{\mathbf{x}_\text{o}} \pi x}\biggl(1-\biggl(1- e^{ - \frac{\eta \beta  x^{\frac{{\alpha_\text{i}}}{2}}}{P} {(\sigma ^2+\gamma)}}\biggr)^{M-L+1}\biggr)\mathrm{d}x.
\end{align}
From the Gamma function $\Gamma(n) = (n-1)!$ for an integer $n$ and the change of variable $\lambda \pi p_i^{\mathbf{x}_\text{o}}x$, we know that for the special case with $D = \infty$ and ${\alpha_\text{i}} = 0$, the two sides of (43) are equal because
\begin{align}
\frac{1}{(\lambda p_i^{\mathbf{x}_\text{o}} \pi)^2}\Gamma(2)=\int_{0}^{\infty} x e^{-\lambda p_i^{\mathbf{x}_\text{o}} \pi x}\mathrm{d}x = \frac{\lambda \pi p_i^{\mathbf{x}_\text{o}}}{2}\int_{0}^{\infty} x^2 e^{-\lambda p_i^{\mathbf{x}_\text{o}} \pi x} \mathrm{d}x =\frac{\lambda \pi p_i^{\mathbf{x}_\text{o}}}{2} \frac{1}{(\lambda p_i^{\mathbf{x}_\text{o}} \pi)^3}\Gamma(3).
\end{align}

In the following, according to (44), we investigate the effect of the parameters $D \neq \infty$ and ${\alpha_\text{i}} \neq 0$ in (43). 

Consider the case $D \neq \infty$ and ${\alpha_\text{i}} = 0$. The ratio of the right over the left integrand functions of (43) is $\frac{2}{{\lambda \pi p_i^{\mathbf{x}_\text{o}}} x}$ that is a decreasing function of $x$. Hence, as $D$ increases from $0$, the ratio $\frac{\int_{0}^{D^2} x e^{-\lambda p_i^{\mathbf{x}_\text{o}} \pi x}\mathrm{d}x}{\frac{\lambda \pi p_i^{\mathbf{x}_\text{o}}}{2}\int_{0}^{D^2} x^2 e^{-\lambda p_i^{\mathbf{x}_\text{o}} \pi x} \mathrm{d}x}$ decreases. Hence, this ratio for the case $D \neq \infty$ is lower bounded by the one for the case $D = \infty$, i.e., $\frac{\int_{0}^{D^2} x e^{-\lambda p_i^{\mathbf{x}_\text{o}} \pi x}\mathrm{d}x}{\frac{\lambda \pi p_i^{\mathbf{x}_\text{o}}}{2}\int_{0}^{D^2} x^2 e^{-\lambda p_i^{\mathbf{x}_\text{o}} \pi x} \mathrm{d}x}\geq \frac{\int_{0}^{\infty} x e^{-\lambda p_i^{\mathbf{x}_\text{o}} \pi x}\mathrm{d}x}{\frac{\lambda \pi p_i^{\mathbf{x}_\text{o}}}{2}\int_{0}^{\infty} x^2 e^{-\lambda p_i^{\mathbf{x}_\text{o}} \pi x} \mathrm{d}x}$, which from (44) leads to 
\begin{align}
\int_{0}^{D^2} x e^{-\lambda p_i^{\mathbf{x}_\text{o}} \pi x}\mathrm{d}x \geq \frac{\lambda \pi p_i^{\mathbf{x}_\text{o}}}{2}\int_{0}^{D^2} x^2 e^{-\lambda p_i^{\mathbf{x}_\text{o}} \pi x} \mathrm{d}x.
\end{align}

Now consider the general case $D \neq \infty$ and ${\alpha_\text{i}} \neq 0$. Here, also consider the term $1-\bigl(1- e^{ - \frac{\eta \beta  x^{\frac{{\alpha_\text{i}}}{2}}}{P} {(\sigma ^2+\gamma)}}\bigr)^{M-L+1}$ (with $M \geq L$) as a weight for the term $x e^{-\lambda p_i^{\mathbf{x}_\text{o}} \pi x}$ in the integrals. As ${\alpha_\text{i}}$ increases from $0$, the ratio $\frac{\int_{0}^{D^2} x e^{-\lambda p_i^{\mathbf{x}_\text{o}} \pi x}\bigl(1-\bigl(1- e^{ - \frac{\eta \beta  x^{\frac{{\alpha_\text{i}}}{2}}}{P} {(\sigma ^2+\gamma)}}\bigr)^{M-L+1}\bigr)\mathrm{d}x}{\frac{\lambda \pi p_i^{\mathbf{x}_\text{o}}}{2}\int_{0}^{D^2} x^2 e^{-\lambda p_i^{\mathbf{x}_\text{o}} \pi x} \bigl(1-\bigl(1- e^{ - \frac{\eta \beta  x^{\frac{{\alpha_\text{i}}}{2}}}{P} {(\sigma ^2+\gamma)}}\bigr)^{M-L+1}\bigr)\mathrm{d}x}$ increases since the ratio of the weight for a small $x$ to the weight for a large $x$ increases. Hence, this ratio for the case ${\alpha_\text{i}} \neq 0$ is lower bounded by the one for the case ${\alpha_\text{i}} = 0$, i.e., $\frac{\int_{0}^{D^2} x e^{-\lambda p_i^{\mathbf{x}_\text{o}} \pi x}\bigl(1-\bigl(1- e^{ - \frac{\eta \beta  x^{\frac{{\alpha_\text{i}}}{2}}}{P} {(\sigma ^2+\gamma)}}\bigr)^{M-L+1}\bigr)\mathrm{d}x}{\frac{\lambda \pi p_i^{\mathbf{x}_\text{o}}}{2}\int_{0}^{D^2} x^2 e^{-\lambda p_i^{\mathbf{x}_\text{o}} \pi x} \bigl(1-\bigl(1- e^{ - \frac{\eta \beta  x^{\frac{{\alpha_\text{i}}}{2}}}{P} {(\sigma ^2+\gamma)}}\bigr)^{M-L+1}\bigr)\mathrm{d}x}\\\geq \frac{\int_{0}^{D^2} x e^{-\lambda p_i^{\mathbf{x}_\text{o}} \pi x}\mathrm{d}x}{\frac{\lambda \pi p_i^{\mathbf{x}_\text{o}}}{2}\int_{0}^{D^2} x^2 e^{-\lambda p_i^{\mathbf{x}_\text{o}} \pi x} \mathrm{d}x}$, which from (45) proves the theorem.

\end{document}